# Decoding Ultrafast Polarization Responses in Lead Halide Perovskites by the Two-Dimensional Optical Kerr Effect

Sebastian F. Maehrlein[1,*], Prakriti P. Joshi[1,*], Lucas Huber[1], Feifan Wang[1], Marie Cherasse[1, †], Yufeng Liu[1], Dominik M. Juraschek[2,3], Edoardo Mosconi[4], Daniele Meggiolaro[4], Filippo de Angelis[4,5,6], X.-Y. Zhu[1,‡]

[1]*Department of Chemistry, Columbia University, USA*
[2]*Harvard John A. Paulson School of Engineering & Applied Sciences, Harvard University, USA*
[3]*Department of Materials Science, ETH Zurich, Switzerland*
[4]*Computational Laboratory for Hybrid/Organic Photovoltaics (CLHYO), CNR-SCITEC, Italy*
[5]*Department of Chemistry, Biology and Biotechnology, University of Perugia, Italy*
[6]*CompuNet, Istituto Italiano di Tecnologia, Genova, Italy*

**The ultrafast polarization response to incident light and ensuing exciton/carrier generation are essential to outstanding optoelectronic properties of lead halide perovskites (LHPs). A large number of mechanistic studies in the LHP field to date have focused on contributions to polarizability from organic cations and the highly polarizable inorganic lattice. For a comprehensive understanding of the ultrafast polarization response, we must additionally account for the nearly instantaneous hyperpolarizability response to the propagating light field itself. While light propagation is pivotal to optoelectronics and photonics, little is known about this in LHPs in the vicinity of the bandgap where stimulated emission, polariton condensation, superfluorescence, and photon recycling may take place. Here we develop two-dimensional optical Kerr effect (2D-OKE) spectroscopy to energetically dissect broadband light propagation and dispersive nonlinear polarization responses in LHPs. In contrast to earlier interpretations, the below-bandgap OKE responses in both hybrid $CH_3NH_3PbBr_3$ and all-inorganic $CsPbBr_3$ perovskites are found to originate from strong hyperpolarizability and highly anisotropic dispersions. In both materials, the nonlinear mixing of anisotropically propagating light fields result in convoluted oscillatory polarization dynamics. Based on a four-wave mixing model, we quantitatively derive dispersion anisotropies, reproduce 2D-OKE frequency correlations, and establish polarization dressed light propagation in single crystal LHPs. Moreover, our findings highlight the importance of**

[*] These authors contributed equally.
[†] Current address: LSI, CEA/DRF/lRAMIS, Ecole Polytechnique, CNRS, lnstitut Polytechnique de Paris, F-91128 Palaiseau, France
[‡] To whom correspondence should be addressed. E-mail: xyzhu@columbia.edu

**distinguishing the often-neglected anisotropic light propagation from underlying coherent quasi-particle responses in various forms of ultrafast spectroscopy.**

**Significance Statement**

Optoelectronic devices, such as photovoltaics or light emitting diodes, are based on the interaction of light with charges in the active material. Lead halide perovskites (LHPs) have emerged as excellent materials for optoelectronics, but mysteries remain as how these materials interact with light. Here, we develop a spectroscopic method to observe the nearly instantaneous responses of LHPs to propagating light fields. This method has allowed us to precisely trace a nonlinear material response resulting from the mixing of different light components overtaking each other during propagation. The resulting oscillatory signals could be easily confused with collective motions of ions or molecules, but are quantitatively accounted by strong nonlinear mixing and polarization dressed light propagation in single crystal LHPs.

**Classification:** Physical Science: Physics

**Introduction**

Understanding the ultrafast polarization response to light fields and the subsequent generation of charge carriers or excitons is key to establishing the photophysical mechanisms in the excellent optoelectronic material system of LHPs (1). The two ionic polarization contributions by the re-orientational motion of organic cations and the deformation of the inorganic cages have been discussed within dynamic screening models (2–4) and large polaron formation(5), respectively and jointly, whereas 3) the immediate electronic polarization response to the light field itself has been neglected so far. In many optoelectronic applications, nevertheless, not only charge carrier transport, but also light propagation right below the bandgap is essential. In LHP nanowire lasers, the lasing modes are known to be red-shifted from excitonic resonances due to efficient coupling to plasmon emission (6). In LHP-based exciton-polariton devices, light-matter coupling red-shifts the hybrid state on the lower polariton branch (7). Propagation of sub-gap light is known to boost the efficiency of LHP photovoltaic cells and light-emitting devices by the so-called "photon recycling" (8). Light propagation strongly influences the function of LHP photonic devices in general (9, 10). A key feature of light propagation near the bandgap is its strong photon energy dependence, as is obvious from the classic Lorentzian model for the dielectric function near an optical resonance (11). However, most photophysical experiments probing carrier/exciton formation, screening, scattering, and nonlinear optical responses employ ultrashort excitation pulses with inherently broad energy distribution and thus convoluted spectral responses. Here, we develop a Fourier-transform based laser spectroscopy technique, 2D-OKE, to investigate light propagation and nonlinear polarization responses directly in the time domain with superior excitation energy resolution near the electronic bandgap.

The 3$^{rd}$ order nonlinear electric polarization $P^{(3)}$ serves as an *in-situ* probe of a material's polarizability and governs the ultrafast macroscopic response to an incident light field. This is employed in a variety of spectroscopies, such as (magneto-) OKE (12, 13), coherent phonon spectroscopy (14, 15), and four-wave mixing (FWM) in general (11). Recently, OKE has been applied to LHP single crystals: Below the bandgap, the dominating non-oscillatory Kerr response of MAPbBr$_3$ (MA = CH$_3$NH$_3$) compared to its all-inorganic counterpart CsPbBr3 was previously attributed to the transient polarization anisotropy caused by liquid-like reorientation dynamics of organic cations (2) and lattice disorder (5). The exponentially decaying responses with above gap excitations were discussed in relation to polaron formation in both materials (5). Interestingly, for excitation energies close to the bandgap in CsPbBr$_3$ at room temperature, time-resolved OKE reveals complex oscillatory features. Such oscillatory transient birefringence signals are usually attributed to coherently excited collective-modes, such as phonons (15–17) or magnons (18, 19), but the strong dependence of the oscillatory frequency on pump photon energy in OKE seems to contradict these origins in LHPs (5). In this work, we unveil a unified source for the Kerr responses in single crystal LHPs by tracing contributions from hyperpolarizability and the peculiar light propagation close to electronic transitions.

**Two-dimensional optical Kerr spectroscopy.** In analogy to other 2D spectroscopy methods (4, 20–26), we use a phase-stable double pump pulse sequence to implement 2D-OKE, **Fig. 1a**. A pulse shaper generates a collinear and identical pump pulse pair ($h\nu_1$ = 2.25 eV, 30 fs pulse duration) with the same polarization, a controllable time delay $\tau$, and precise phase difference modulated shot by shot (27); see **Materials and Methods**. We additionally implement a standard 4-sequence phase cycling scheme to discriminate against incoherent absorption and scattering background, and a rotating frame to increase the pump frequency resolution while maintaining rapid data collection (20, 26, 27) (see **Materials and Methods,** and SI Appendix, 1.2-1.5). Fourier transform with respect to $\tau$ provides precise pump-energy resolution; particularly interesting near electronic resonances where highly dispersive linear and nonlinear optical responses are expected. A successive below-bandgap probe pulse ($h\nu_2$ = 1.55 eV, 45 fs pulse duration), delayed by time *t* with respect to the later pump pulse, monitors the anisotropic 3rd order polarization response by a transient birefringence $S(t,\tau)$. A balanced detection further discriminates against isotropic transmission changes, e.g. transient absorption. We use solution grown macroscopic single crystal $CsPbBr_3$ and $MAPbBr_3$ (see **Materials and Methods**, and SI Appendix, 1.1 & Fig. S1) with thicknesses in the range of 50 - 500 μm to study light propagation macroscopically with respect to a defined crystallographic structure.

**Fig. 1b** illustrates how time resolved OKE probes light propagation through a dispersive crystal with birefringence characterized by an anisotropic refractive index, $\Delta n_0(\nu)$, as is the case in the orthorhombic phase of $CsPbBr_3$ or $MAPbBr_3$. The pump pulse (red) entering the crystal decomposes into two orthogonal fields with fast (blue) and slow (orange) propagating velocities. The two orthogonal pump polarizations and the time-delayed probe pulse nonlinearly mix to yield a 3rd order nonlinear polarization and thus a contribution to the OKE signal. In general, this corresponds to a collinear, partly degenerate and polarization-resolved FWM process (11, 28). Additionally, the highly dispersive nature near the bandgap is reflected in the pump-probe walk-off due to velocity mismatching and in the temporal spread of the pump pulses due to group velocity dispersion (GVD) (29).

**RESULTS**

Before presenting 2D-OKE results, we establish the complexity of single pump-pulse (1D) OKE experiments with improved signal-to-noise ratio by using a balanced detection. The 1D-OKE response from $CsPbBr_3$ at room temperature with close-to-gap excitation ($h\nu_1$ = 2.25 eV), **Fig. 1c**, shows a clear oscillatory signal. The initial high frequency response of 5.3 THz is beyond the frequency range of any known lattice mode of this crystal (see our DFT calculations in SI Appendix, Fig. S13). We can further exclude stress or photodoping effects that are unable to increase a phonon mode to ~5.3 THz (SI Appendix 2.3, Figs. S12 and S14). Moreover, a time frequency analysis, **Fig. 1d**, reveals a stark frequency jump from ~5.3 THz (region 1) to ~2.5 THz (region 2) at ~ 1.5 ps. Our simulations show (SI Appendix, 2.2) that such strong phonon-phonon coupling (30, 31) would require parametric down conversion (32) with unprecedentedly large nonlinear coupling coefficients as compared to those known in other solids (31). The high frequencies ~5.3 THz is also not possible from changes to the phonon frequency in the presence of photo-doped carriers (SI Appendix, 2.3).

**2D-OKE in CsPbBr$_3$.** We now turn to 2D-OKE to decode these transient birefringence signals and to establish polarization-dressed light propagation near the bandgap as the unified origin of the ultrafast Kerr response in CsPbBr$_3$ and CH$_3$NH$_3$PbBr$_3$. **Figs. 2a** and **2c** shows the excitation energy (hv$_p$) dependent transient birefringence as a function of pump-probe delay $t$ for CsPbBr$_3$ at T = 295 K and 92 K, respectively. The energy-time domain 2D OKE signal (pseudo color) is obtained by the real part of the signal's Fourier transform (FT) with respect to the pump-pump delay $\tau$ (see SI Appendix, Fig. S2 for time-time-domain data). Prominently, the signal exhibits strong oscillatory features decreasing in frequency with increasing hv$_p$.

The 2D-OKE in **Fig. 2a** or **2c** exhibits an underlying fan-like superstructure (region 1), showing the strong dependence of modulation frequency (v$_{OKE}$) on pump photon energy (hv$_p$). We quantify the anti-correlation by carrying out a second Fourier transform with respect to the pump-probe delay $t$. At 295 K, **Fig. 2b**, the anti-correlation between v$_{OKE}$ and hv$_p$ is represented by one dominant branch while at the lower temperature of 92 K, **Fig. 2d**, multiple branches with different slopes are resolved. As we show below in our modeling, the strong anti-correlations of v$_{OKE}$-hv$_p$ are caused by the hyperpolarizability experienced by light propagating in the highly dispersive and anisotropic region near the bandgap. A particularly revealing feature in the 2D-OKE spectra (**Fig. 2a** or **2c**) is the sudden stop of the fan-like oscillatory superstructure with a bump-like feature (region 2), followed by long-lived low frequency oscillation (region 3). Such sudden change in oscillatory frequency is unphysical for coherent phonons, but is consistent with the dispersive propagation as detailed below. Supporting this assignment of light propagation, we find that the pump-probe delay $t_1(v_p)$ (blue markers), when region 2 occurs is proportional to the thickness of the crystal (SI Appendix, Figs. S7 and S8). Affirming the reliability of the 2D-OKE method based on projection slice theorem(20), we show in insets above each panel in **Fig. 2** that the Excitation energy-integrated 2D-OKE traces (red) are in good agreement with the single pulse excitation 1D-OKE (gray/blue).

**2D-OKE in MAPbBr$_3$.** The CsPbBr$_3$ crystal at either 295 K or 92 K is in the orthorhombic phase, which is anisotropic and thus birefringent(33). We find that the lack of oscillatory OKE signatures in the hybrid CH$_3$NH$_3$PbBr$_3$ is not related to dynamic disorder nor organic cation orientation degrees of freedom, but due to the crystallographic phase. MAPbBr$_3$ is in the non-birefringent cubic phase at room temperature (295 K) and the birefringent orthorhombic phase below 150 K(34). In the room temperature cubic phase, **Fig. 3a**, we reveal the evolution of the strong quasi-static Kerr response, which broadens towards the electronic band gap (E$_g$ ~2.32 eV at 295 K); fully consistent with earlier OKE studies(5). The signal vanishes towards the bandgap due to a drastic decrease in penetration depth and therefore decreased effective probing volume. The 2D-OKE also unveils an abrupt decay of the Kerr signal, which is averaged out and thus hidden in conventional 1D-OKE experiments due to the lack of energy selectivity (Fig. **3a,** upper inset). Fourier transform with respect to pump-probe delay *t* shows only a broad zero frequency feature (Fig. **3b**). In stark contrast to results at room temperature, 2D-OKE in the low temperature orthorhombic phase (*Pnma*) of MAPbBr$_3$, **Fig. 3c**, shows a strong oscillatory signal, which is similar to responses in the orthorhombic phase of CsPbBr$_3$ in **Fig. 2**. Fourier transform of the 2D-OKE with respect to the pump-probe delay *t*, **Fig. 3d**, again reveals the anti-correlation between ν$_{OKE}$ and hν$_{ex}$ in two apparent branches for MAPbBr$_3$ in the orthorhombic phase, with ν$_{OKE}$ spanning one order of magnitude, from ~ 1 THz to ~ 10 THz.

**Model.** For both CsPbBr$_3$ and MAPbBr$_3$ in the orthorhombic phase, the exceptionally large spread in frequency ν$_{OKE}$ and the abrupt decay at $t_1$ are unphysical for coherent phonons or other collective modes. As suggested in early FWM studies, light propagation can strongly affect FWM signals and especially 2D spectroscopy line shapes in isotropic media such as dense gases (35, 36), solutions (37) or cubic semiconductors (38). In LHPs, anisotropic light propagation seems to play a crucial role. Therefore, we consider an instantaneous electronic polarization (via hyperpolarizability) and propagating light fields in single crystals by taking into account the influence of spectral dispersion, static birefringence and nonlinear mixing via the 3$^{rd}$ order nonlinear susceptibility $\chi^{(3)}$. We trace the propagation of all three interacting light fields that mix via $\chi^{(3)}$ to produce new polarization components:

$$P_j^{(3)}(t,z) = \epsilon_0 \int_{-\infty}^{t} dt' \int_{-\infty}^{t'} dt'' \int_{-\infty}^{t''} dt''' \; \chi_{jklm}^{(3)}(t,t',t'',t''',z) \; E_k^{\text{p}}(t',z) \; E_l^{\text{p}}(t'',z) \; E_m^{\text{pr}}(t''',z) \quad (1)$$

with probe field $E^{pr}$, pump fields components $E^p$, vacuum permittivity $\epsilon_0$ and collinear propagation along $z$. To model the instantaneous Kerr response arising from the hyperpolarizability, we set $\chi^{(3)}_{jklm}(t, t', t'', t''', z)$ to a product of $\delta$-functions in time $\sim \chi^{(3)}_{jklm} \delta(t - t') \delta(t' - t'') \delta(t'' - t''')$ and thus independent of light frequencies, whereas the equilibrium linear refractive index tensor $n_0(\nu)$ exhibits a strong dispersion near the bandgap and static birefringence $\Delta n_0(\nu)$ in the orthorhombic phase (see SI Appendix, 1.6). This includes pump-probe walk-off from group velocity mismatch (group refractive indices: $n_g^p(\nu) > n_g^{pr}$) and the temporal spread of the pump pulses, both due to GVD.

**Isotropic cubic phase.** For cubic (*Pm3m*) MAPbBr$_3$ at room temperature with $\boldsymbol{E}^{pr}$ parallel to a crystal axis $\hat{\boldsymbol{y}}$, we find that the dominating contribution is the nonlinear mixing term of a single pump pulse with itself via $\chi^{(3)}_{xxyy} E_x^p E_y^{*p}$ while propagating through the depth of the sample, shown in false colors in Fig. **4a**. We calculate the signal fields $E_x^s(t, z)$ emitted by the local nonlinear polarizations $P_x^{(3)}(t, z)$ along the probe field trajectory (green lines) for distinct pump-probe delays $t$. Additionally, we account for the propagation of each signal field and for the projection of the accumulated signal given by the balanced detection geometry after the sample. Adding a second pump pulse with delay $\tau$ to this simulation (including phase cycling and rotating frame) allows for the modeling of the 2D OKE signal(39). Strikingly in Fig. **4b**, this simulation accounting solely for instantaneous polarization responses reproduces all features of the experimental OKE response in the cubic phase MAPbBr$_3$ in **Fig. 3a**: The fan-like OKE response towards the bandgap (region 1) results from the strong GVD and walk-off of each individual nonlinear polarization component, both resolved with an unprecedented meV excitation energy resolution by the 2D-OKE method. The pronounced bump-like feature (region 2) stems from the interference of the pump pulse with itself at the backside of the sample (see Fig. **4a**). The 2D frequency map in Fig. **4c** is dominated by a broadened zero-frequency feature, in agreement with the experimental frequency map in Fig. **2c**.

**Birefringent orthorhombic phase.** For CsPbBr$_3$ and MAPbBr$_3$ in the *Pnma* orthorhombic phase, the key finding is the strong oscillatory signature, which fans out towards the bandgap. In the birefringent crystal, we consider pump field projections on fast and slow axes with refractive indices $n_x$ and $n_y$, respectively; the birefringence is quantified as $\Delta n_0(\nu) = n_x(\nu) - n_y(\nu)$. Thus, the corresponding projections $E_x^p$ and $E_y^p$ propagate with different velocities through the crystal while nonlinearly mixing via $\chi_{xxyy}^{(3)} E_x^p E_y^{*p}$. The simulation results in Fig. **4d** unveil a spatiotemporally oscillating nonlinear mixing term in addition to the effects discussed above for the cubic phase. Thus, the probe pulse experiences an oscillating longitudinal transient phase grating(28, 40, 41) produced by dispersively co-propagating orthogonal pump pulse projections. By further including the static birefringence of the probe and signal fields ($E^{pr}$ rotated from $\hat{y}$ by angle $\phi$), by a global fit of the refractive index dispersion $n(\nu)$ from $t_1$ and by considering a FWM phase mismatch $\Delta k$ (see below and Ref. (39)), the model result in Fig. **4e** reproduces the complex 2D-OKE signal of the orthorhombic phase. Thus, we fully decoded the sophisticated Kerr signals in bulk LHPs: The initial high-frequency oscillatory signal (region 1) is due to the instantaneous nonlinear polarization from propagating pump pulse projections mixing with each other (via off-diagonal $\chi^{(3)}$), traced by a co-propagating probe pulse. The intermediate frequency region 2 stems from the additional spatiotemporal pump interference pattern at the backside of the sample, producing a bump-like structure between the latest possible probe overlap with the fast and slow pump component (dashed white lines from experimental data in Fig. **2c**). Lastly, the long-lived low-frequency signal (see Figs. **1c**, and **2a**) is caused by the same effects of the counter-propagating pump reflection (region 3).

In an analytic model, given by Huber et al.(39), we can identify the main modulation frequency branches by considering the FWM phase mismatch $\Delta k^\pm = k_3^{pr} - k_4^{signal} \pm (k_2^p - k_1^p)$, where '$\pm$' can be seen as a positive and negative sideband generation from $E_1^{*p} E_2^p$ and $E_1^p E_2^{p*}$, respectively. Formally, these two contributions correspond to the rephasing and non-rephasing signal(21, 27). From $\Delta k^\pm$, we can estimate the four modulation frequency branches(39)

$$f_\mp^\pm \approx \frac{1}{2\pi} \frac{\Delta k^\pm}{t_{1,\mp}} d = \frac{\nu^{pr} \Delta n_0(\nu_{pr}) \pm \nu^p \Delta n_0(\nu_p)}{n_G(\nu_p) \mp n_G(\nu_{pr})}, \quad (2)$$

where $\Delta n_0(\nu)$ is again the static birefringence, $n_G(\nu)$ is the group refractive index, and $t_{1,\mp} = d(n_G(\nu_p) \mp n_G(\nu_{pr}))/c$ corresponds to the latest overlap of the probe pulse with the co- ($t_{1,-}$) and counter- ($t_{1,+}$) propagating pump fields at the backside of the sample $z = d$. Thus in Fig **4f**, $f_-^\pm$ and $f_+^\pm$ are the two frequency branches for co- (red & blue dashed line) and counter- (gray dashed lines) propagating pump and probe pulses, respectively. Plotting these analytically predicted branches onto the experimental data in Fig. **2d** (dashed black lines), confirms the striking agreement and comprehensive consistency of our model. The weaker side branches in the experimental data (Fig. **2d**, gray dashed lines) are exact replica of the main features shifted up and down along the excitation energy axis by the amount of their modulation energy $hf_\mp^\pm$.

**DISCUSSION**

The agreement between experimental data of both materials (hybrid and inorganic) in both phases (orthorhombic and cubic) and the modeling solely based on a temporally $\delta$-like 3$^{rd}$ order polarization response function, points towards a unified origin of complex Kerr-responses below and close to the optical bandgap. Polarization contributions from inorganic lattice, cation reorientation and hybrid modes would require slightly retarded response functions depending on inertia and resonance frequencies of these structural dynamics. Here, such underlying influences of the ionic polarizabilities cannot be ruled out, but are found to be overwhelmed by hyperpolarizability contributions. Hence, previous OKE interpretations of dominating orientational contributions in hybrid LHPs and inorganic lattice contributions should be carefully reconsidered depending on their excitation proximity to the optical bandgap and sample thickness (2, 5). Our finding of the dominating hyperpolarizability suggests the picture of intense light fields travelling with nonlinear polarization clouds in the near bandgap region in LHPs.

For comparison, we carry out studies on two polar semiconductors with similar band gaps, the isotropic ZnTe and the birefringent GaSe, both prototype nonlinear optic crystals(42). Whereas the quasi-static 2D-OKE signal of ZnTe compares well to the nonlinear response of cubic MAPbBr$_3$, hints of the oscillatory cross polarized $\chi^{(3)}$ mixing effects in GaSe only occur under very specific phase matching angles and with much weaker signal to noise ratio (SI Appendix, Fig. S6). Thus, our results confirming a substantial $\chi^{(3)}$ in LHPs support previously suggested applications for nonlinear photonics (43, 44).

Further, we establish 2D-OKE as a quantitative probe of the highly dispersive and anisotropic light propagation in the vicinity of electronic transitions. Based on **Eq. 2**, we directly extract absolute values for the group index mismatch $\Delta n_G(\nu)$, Fig. **5a**, and dispersive birefringence $\Delta n_0(\nu)$, Fig. **5b**, with a precision of 10$^{-3}$ in the latter and a wavelength resolution of ~ 0.5 nm (39). We reveal a doubling of the birefringence $\Delta n_0(\nu)$ by cooling to 92 K, indicating anisotropic structural changes within the orthorhombic phase. Thus, we trace how anisotropic light propagation is strongly influenced by position and steepness of the optical bandgap. This can be seen from different dispersion curvatures of the birefringence and group index mismatch at 92 K compared to 295 K, even in the same structural phase (**Fig. 5**). Generally, this precise knowledge of polarization dependent group velocities and birefringence is essential to dispersion engineering of polaritons and their degeneracy in Bragg cavities(7, 45) or the design of perovskite nanocrystal lasers (6, 46) and photonic devices (10, 44).

In conclusion, our results provide a unified picture of Kerr-responses near the bandgap of semiconductors in general and LHPs in particular. The 2D method provides meV energy-resolution and fs time-resolution in OKE responses; the resulting 2D-OKE spectral signatures generally contain information about dispersive birefringence, group velocity dispersion, and bandgap absorption. This technique may allow us to distinguish transient birefringence signals of phonons, spin dynamics, or other collective modes, from the nonlinear light propagation phenomena in various fields of ultrafast spectroscopy. In the case of LHPs, we find the stark contrast in the Kerr signals of $CsPbBr_3$ and $MAPbBr_3$ at room temperature to be dominated by polarization dressed light propagation in different crystallographic phases rather than by structural dynamics (2, 5). Our findings represent a bulk analogy to polarization beats in optical fiber communication, which are used for fast multiplexing via high bandwidth polarization switching (47) and LHPs may find applications in integrated photonic devices for polarization based light modulation. The precisely determined dispersion anisotropy can be additionally used for dispersion engineering of polarization-split polaritons in LHP cavities (7). Understanding Kerr polarization states in LHPs might further pave the way to studies on non-classical states of light in this material class (48).

**MATERIALS AND METHODS**

**Sample preparation.** For the synthesis of $CsPbBr_3$ single crystals, the precursor solution (0.38 M) was prepared by mixing cesium bromide (CsBr, Aldrich, 99.999%) and lead bromide ($PbBr_2$, Aldrich, ≥98%) with molar ratio of 1:1 in dimethylsulfoxide (DMSO, EMD Millipore Co., anhydrous ≥99.8%). After being fully dissolved, the solution was titrated with methanol (Alfa, 99.9%) until the yellow-orange precipitate did not re-dissolve. Then it was further stirred at 50 °C until a yellow permanent suspension formed. Before the crystal growth, the precursor solution was filtered by a PTFE filter with 0.22 μm pores. Methanol was used as the anti-solvent for slow vapor diffusion and the crystal growth was quenched when the desired morphology was achieved(49). A similar method was used for the preparation of MAPbBr$_3$ single crystals(50). The precursor solution (0.45 M) was formed by adding methylammonium bromide (MABr, Dyesol, 98%) and PbBr$_2$ with 1:1 molar ratio into N,N-dimethylformamide (DMF, Aldrich, anhydrous 99.8%). A mixture of dichloromethane (CH$_2$Cl$_2$, Aldrich, ≥99.5%) and nitromethane (Aldrich, ≥96%) was used as the anti-solvent for MAPbBr$_3$ crystal growth. All solid reactants were dehydrated in a vacuum oven at 150 °C overnight and all solvents were used without further purification.

**Experimental setup.** Ultrashort laser pulses (energy of 0.8 mJ, center wavelength 800 nm, duration of 30 fs, and a repetition rate 10 kHz) from a Ti:sapphire regenerative amplifier were split to serve as pump and probe pulses. The pump was frequency-converted by a homebuilt NOPA with a center wavelength of 550 nm (55nm full width), see e.g. SI Appendix, Fig. S3. The excitation energy incident on the sample was $35 \pm 5$ nJ per pump pulse pair (160 μm spot diameter) and the active dispersion compensation (via pulse shaping) optimized to achieve pump pulse durations of 30 fs. The probe beam was focused to 30 μm diameter on the sample under normal incidence (pulse energy < 0.2 nJ). For 1D-OKE in Figs. 1c,d, S9 and S10, the pulse shaper was bypassed, and the NOPA was optimized for more narrow band spectra with central wavelength of 550 nm (pulse duration < 30 fs, full spectral widths of < 25 nm). See details in SI Appendix 1.2.

**Pulse shaping and phase cycling.** We generate double pulses from the NOPA output pulse by using an AOM based visible pulse shaper (PhaseTech QuickShape Visible) to control the pump-pump delay $\tau$ and their relative phase on a shot-to-shot basis (at 10 kHz). We use a common 4-sequence *phase cycling* scheme of $S(0,0) - S(0,\pi) + S(\pi,\pi) - S(\pi,0)$, with $S(\phi_1,\phi_2)$ being the detected signal excited by a pump pulses pair with phases $\phi_1$ and $\phi_2$, respectively(20, 27). We add a *rotating frame* with frequency $\omega_F$ by an additional phase factor $e^{-i\tau\omega_F}$ (e.g. $\omega_F/2\pi =$ 500 THz), depending on the spectral region of interest. Both *rotating frame* and *phase cycling* is implemented by the phase shaper software (PhaseTech QuickControl). We exclude artifacts from both by performing single pulse excitation (1D-OKE) reference measurements (insets in Figs. 2 and 3) with identical beam path by using the phase shaper only as a 5 kHz chopper. The pump-probe delay $t$ is introduced by a conventional mechanical delay stage. See details in SI Appendix 1.3.

**Balanced detection and data acquisition.** The balanced detection consists of half-wave plate and Wollaston prism followed by a focusing lens and two identical fast Si-photodiodes. The relative difference of the respective intensities $\Delta I/I_0$ scales linearly with rotation of the probe polarization for small angles. The single shot difference signal $\Delta I(t, \tau, \phi_1, \phi_2)$ is digitized and acquired at a 10kHz rate in conjunction with the corresponding pump-pump-pulse sequence. See details in SI Appendix 1.5.

**Model.** We simulate $P_j^{(3)}(t,z)$ according to the FWM equation shown in the main text. We assume that the fields are not depleted by the nonlinear process(11) and that the polarization components $P_j^{(3)}(t,z)$ emit a nonlinear signal field $E^{NL}(t,z)$ at every slice $z$, which then co-propagates with the probe field. The transmitted probe field $E^{pr}$ and integrated nonlinear field $E^{NL}$ is then projected on two polarization detection directions given by a half-wave plate and a Wollaston prism via Jones calculus (with balanced static signal). For normal incidence on the (101) crystal surface, the orthorhombic space group *Pnma* of CsPbBr$_3$ permits Kerr signals from $\chi_{xxxx}^{(3)}, \chi_{yyyy}^{(3)}, \chi_{xxyy}^{(3)} = \chi_{xyyx}^{(3)} = \chi_{xyxy}^{(3)}$ and $\chi_{yyxx}^{(3)} = \chi_{yxxy}^{(3)} = \chi_{yxyx}^{(3)}$ (Ref.[1]). The cubic MAPbBr$_3$ at room temperature has space group *Pm3m* and allows for coupling via $\chi_{xxxx}^{(3)} = \chi_{yyyy}^{(3)}$ and $\chi_{xxyy}^{(3)} = \chi_{xyyx}^{(3)} = \chi_{xyxy}^{(3)} = \chi_{yyxx}^{(3)} = \chi_{yxxy}^{(3)} = \chi_{yxyx}^{(3)}$. All allowed tensor elements were assumed to be of same magnitude (in arb. u.). See details in SI Appendix 1.6 and Ref. (39).

## ACKNOWLEDGMENTS AND FUNDING SOURCES

We thank Louis Brus, Simon Billinge, Chris Middleton, and Nicola Spaldin for fruitful discussions, Kiyoshi Miyata for the initial experimental steps towards 2D-OKE. XYZ acknowledges support for the 2D experiments by the Vannevar Bush Faculty Fellowship through Office of Naval Research Grant # N00014-18-1-2080 and the modeling by the US Department of Energy, Office of Energy Science, grant DE-SC0010692. The design and construction of the 2D-OKE spectrometer was supported in part by the US Air Force Office of Scientific Research (AFOSR) grant FA9550-18-1-0020 and by the Columbia Nano Initiative. SFM was supported by a Feodor Lynen Fellowship of the Alexander von Humboldt Foundation. LH was supported by the Swiss National Science Foundation under project ID 187996. MC acknowledges support from the Alliance Program between Columbia University and Ecole Polytechnique. DMJ was supported by the Swiss National Science Foundation (SNSF) under Project ID 184259.

## AUTHOR CONTRIBUTIONS

XYZ, SFM, and PPJ conceived and initiated this work. SFM designed and developed new experimental methods. SFM, PPJ, FW, LH, MC and YL carried out the experimental measurements. LH developed model and carried out 2D-OKE simulations. FW grew the crystals. DMJ, FdA, EM and DM carried out phonon simulation. SFM, LH, FW and PPJ carried out data analysis. SFM and XYZ wrote the manuscript. All authors read and commented on the manuscript.

**COMPETING INTEREST STATEMENT**

The authors declare no conflict of interest.

**SI APPENDIX**

Detailed methods and extended data.

**FIGURES**

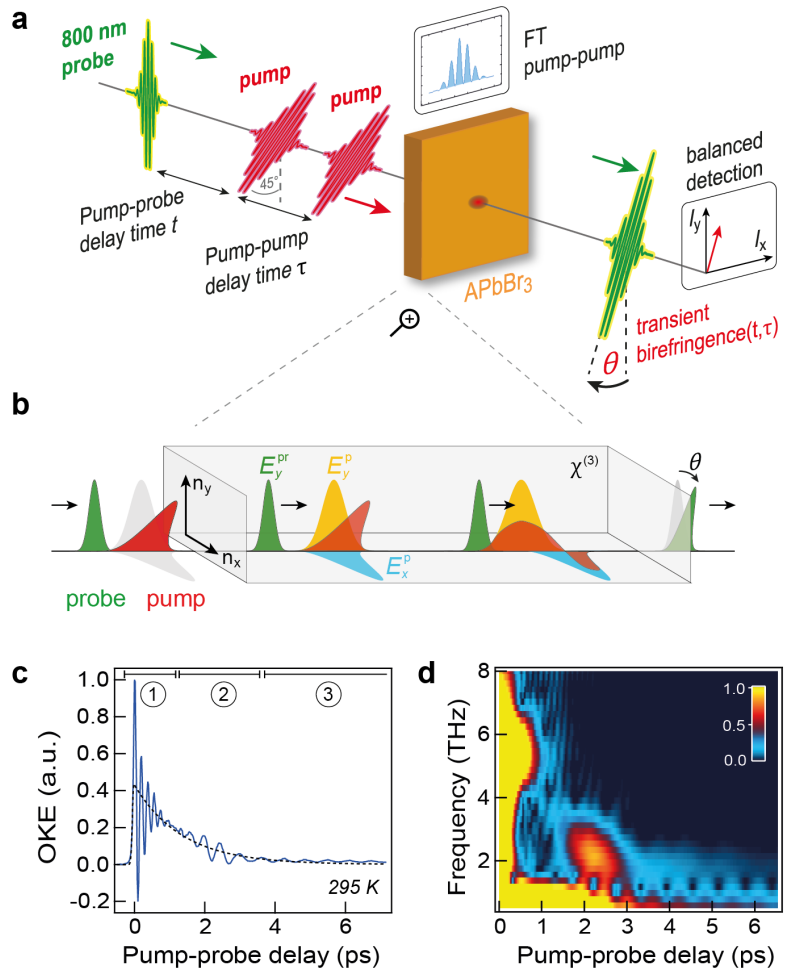

**Figure 1|Experimental scheme and 1D OKE results. a**, Experimental Scheme 2D-OKE: Two phase-controlled, co-propagating pump pulses induce a nonlinear polarization response while interacting with a time-delayed probe pulse; traced via transient birefringence $S(t,\tau)$ as a function of pump-pump $\tau$ and pump-probe delay $t$. **b**, $S(t,\tau)$ precisely monitors anisotropy and dispersion of light propagation, dominated by nonlinear $\chi^{(3)}$ mixing of perpendicular pump field projections $E_x^p$ and $E_y^p$ while interacting with - and propagating with respect to - the probe field $E_y^{pr}$. **c**, Conventional ultrafast OKE in CsPbBr$_3$ at 295 K. Single pulse (1D) excitation ($h\nu_p = 2.25$ eV) exposes polarization dynamics with three distinct frequency regions (circle numbers) in the vicinity of the optical band gap. **d**, Pseudo color (FT amplitude) representation from a moving-window Fourier analysis of the OKE signal in **c**, unveiling a highly nonlinear frequency evolution.

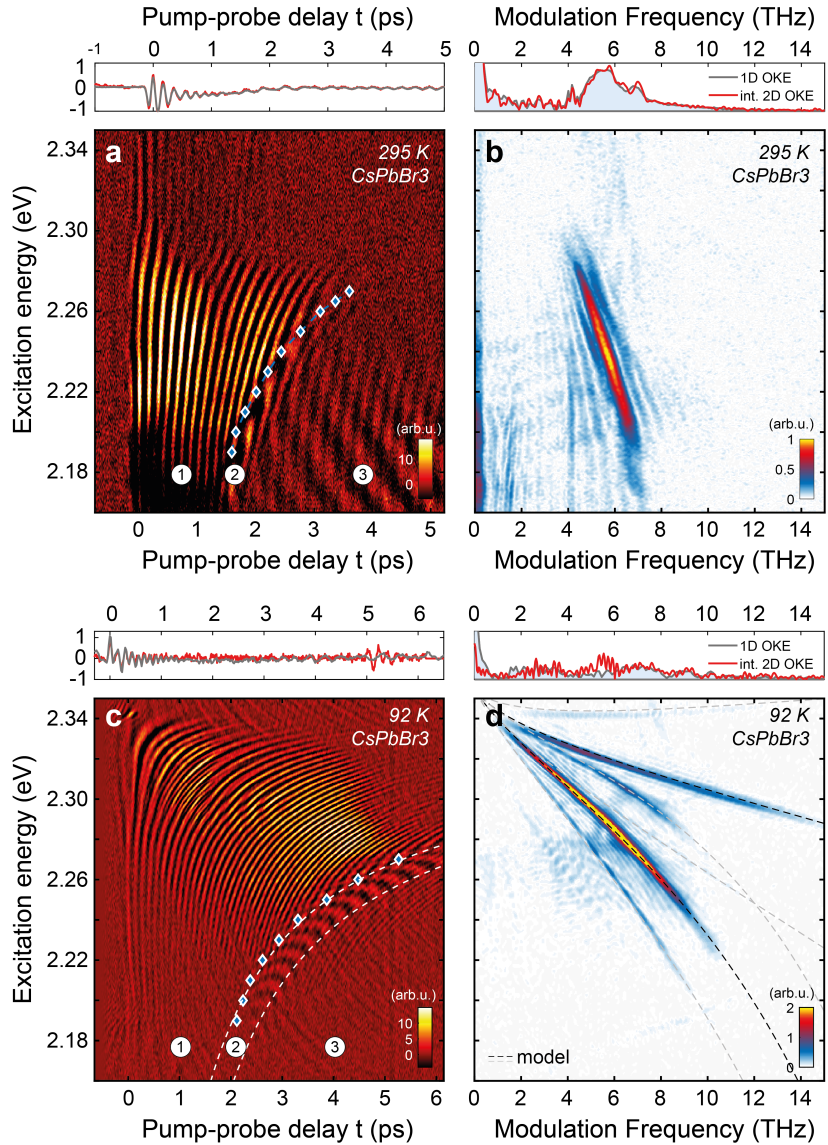

**Figure 2 | 2D OKE in CsPbBr₃.** Energy-time domain 2D-OKE signal (pseudo color) at 295 K given by the FT of the transient birefringence $S(t,\tau)$ with respect to the pump-pump delay $\tau$. Consistent with 1D-OKE, three distinct regions can be identified (circle numbers). End of region 1 ($t_1$) is determined in the phase map and shown as blue markers. **b**, 2D FT amplitude: Correlations of excitation energy $h\nu_p$ and modulation frequency $\nu_{OKE}$, derived from **a** by FT with respect to pump-probe delay $t$. **c**, **d**, Same as **a**, **b**, but at 92 K. **c,** Dashed lines are fits to determine the refractive and group index from the 2D-OKE fingerprint. **d**, Multiple frequency-frequency correlation branches in full agreement with our model (dashed lines): Two main branches defined by FWM phase matching (black) and higher-order side bands (gray). All upper insets: Agreement of energy integrated 2D-OKE and single pulse 1D-OKE under identical experimental conditions.

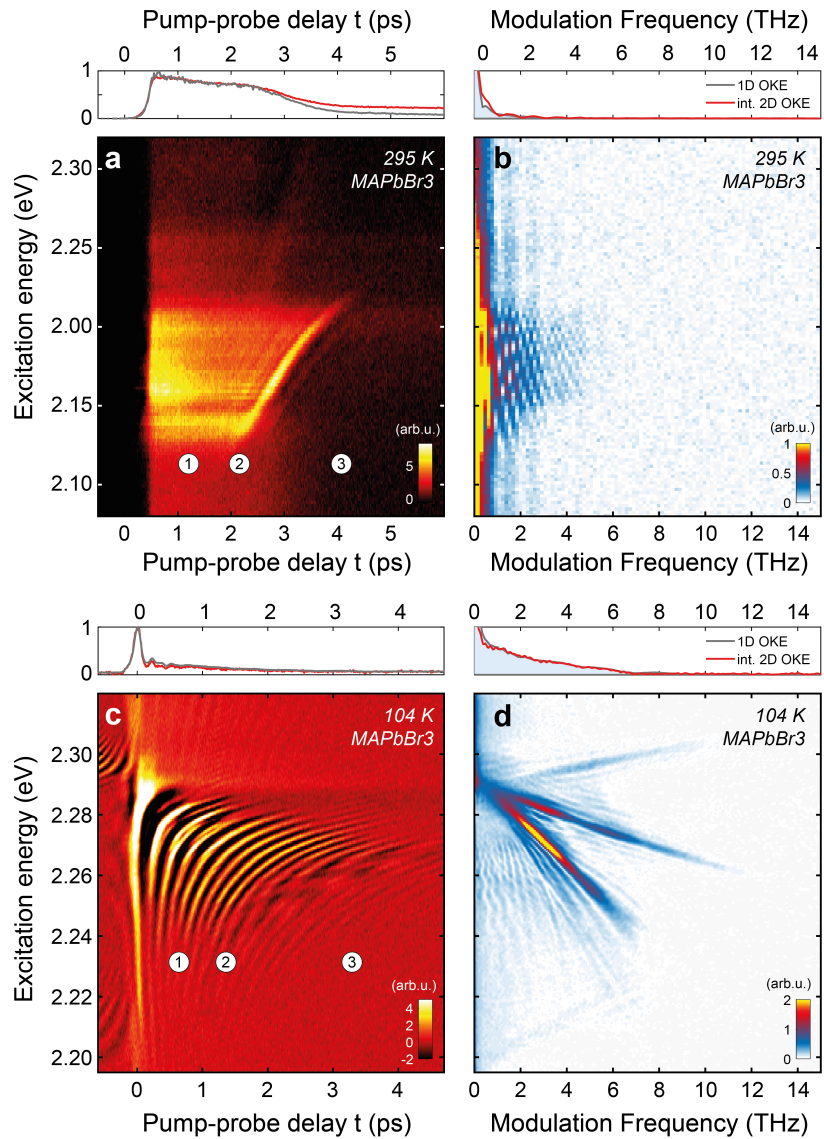

**Figure 3 | 2D OKE in MAPbBr$_3$. a**, Energy-time domain and **b**, Energy-frequency domain of the 2D OKE signal in cubic phase MAPbBr$_3$ at 295 K, analogous to Fig. 2. **c, d**, Same as **a,b**, but in the orthorhombic phase at 104 K. Clear similarity to orthorhombic CsPbBr$_3$ unveils structural phase and its birefringence properties as main source for the stark contrast between MAPbBr$_3$ and CsPbBr$_3$ OKE at room temperature.

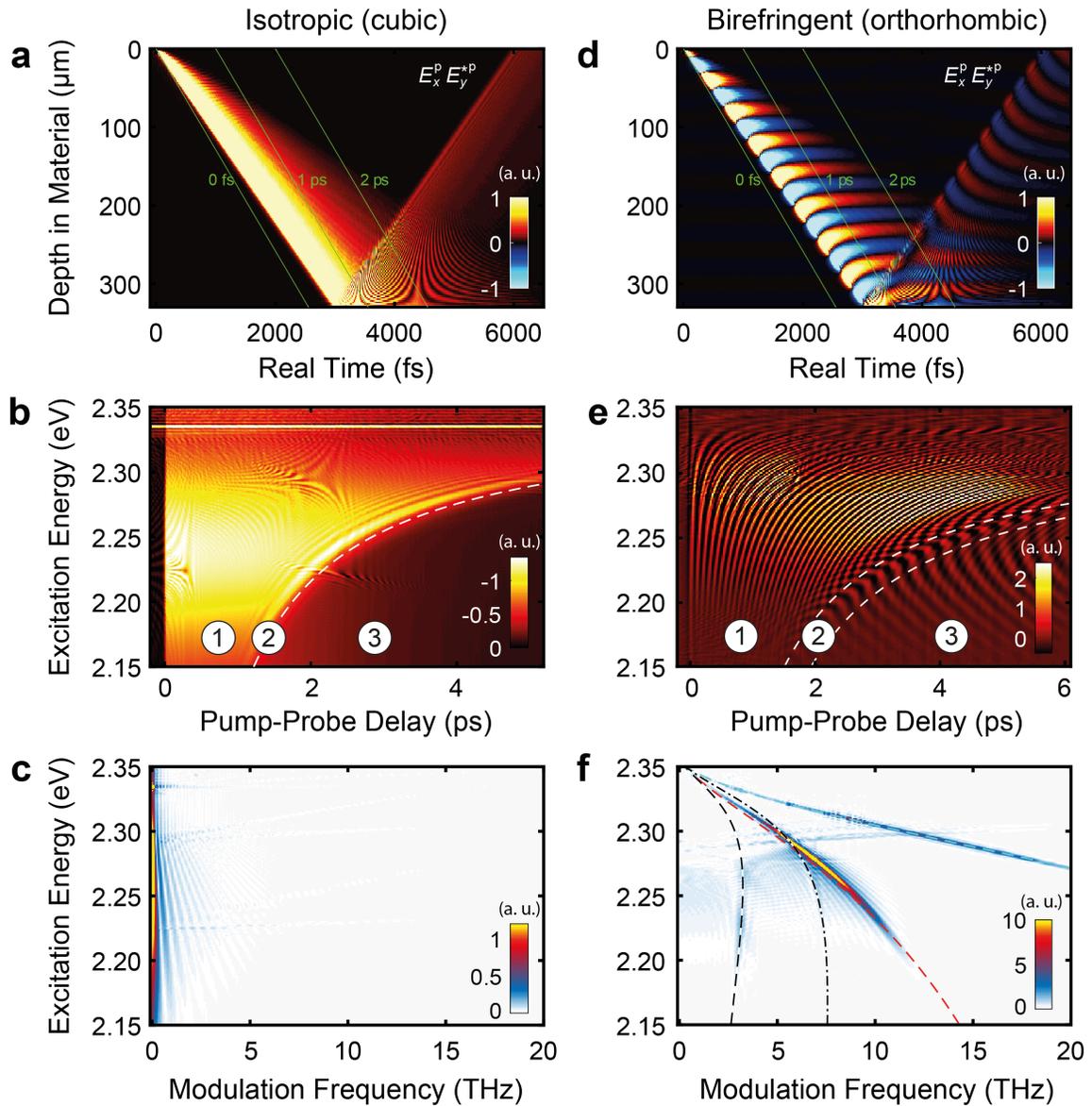

**Figure 4 | Polarization sensitive four-wave mixing simulation. a**, Spatiotemporal calculation of the instantaneous nonlinear mixing term $\chi^{(3)}_{xxyy}E^p_x(t',z)\ E^{*p}_y(t',z)$ (false colors) for the isotropic cubic phase. Fields propagate in space $z$ (top to bottom) and real time $t'$ (x-axis). At the end of the sample (320 μm), they get partially reflected and propagate in opposite direction. Probe pulse trajectories are shown as green lines. Pump pulse GVD and pump-probe walk-off are clearly resolved by different slopes. **b**, Corresponding simulated 2D-OKE signal in full analogy to the experiments; including double pulse excitation, integrated nonlinear mixing term along each probe pulse trajectory, and consecutive signal propagation and projection by balanced detection. **c**, 2D-OKE Frequency map from FT of b. **d, e, f**, Analog to **a, b, c**, but for the orthorhombic phase (thickness $d$=480 μm for **e,f** ; probe polarization $\phi = 0°$). **e**, Model reproduces experimental data, including precise region 2 boundaries from Fig. 2c (white dashed lines) **f**, 2D-frequency correlations perfectly fit experimental data in Fig. 2d and follow expected FWM phase matching conditions for co- (red and blue dashed lines) and counter- (black broken lines) propagating pump fields in **f**.

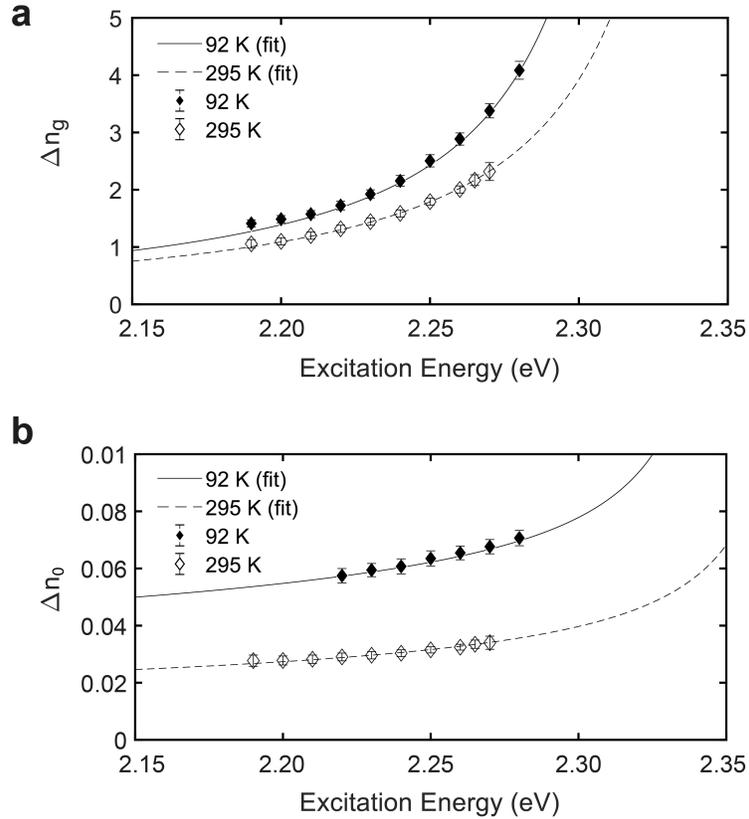

**Figure 5 | Derived dispersion anisotropy. a**, Dispersion of the group index mismatch between pump and probe pulse $\Delta n_g(\nu)$ and **b**, Dispersive pump birefringence $\Delta n_0(\nu)$ in the vicinity of the optical band gap. Both derived from 2D OKE fingerprints in **Fig. 2a,b** and **Fig. 2c,d** for 295 K and 92 K, respectively.

# Decoding Ultrafast Polarization Responses in Lead Halide Perovskites by the Two-Dimensional Optical Kerr Effect
## --- Supporting Information Appendix ---

Sebastian F. Maehrlein[1,*], Prakriti P. Joshi[1,*], Lucas Huber[1], Feifan Wang[1], Marie Cherasse[1,†], Yufeng Liu[1], Dominik M. Juraschek[2,3], Edoardo Mosconi[4], Daniele Meggiolaro[4], Filippo De Angelis[4,5,6], X.-Y. Zhu[1,‡]

[1]*Department of Chemistry, Columbia University, USA*
[2]*Harvard John A. Paulson School of Engineering & Applied Sciences, Harvard University, USA*
[3]*Department of Materials Science, ETH Zurich, Switzerland*
[4]*Computational Laboratory for Hybrid/Organic Photovoltaics (CLHYO), CNR-SCITEC, Italy*
[5]*Department of Chemistry, Biology and Biotechnology, University of Perugia, Italy*
[6]*CompuNet, Istituto Italiano di Tecnologia, Genova, Italy*

# 1. Methods
### 1.1. Sample preparation

For the synthesis of $CsPbBr_3$ single crystals, the precursor solution (0.38 M) was prepared by mixing cesium bromide (CsBr, Aldrich, 99.999%) and lead bromide ($PbBr_2$, Aldrich, ≥98%) with molar ratio of 1:1 in dimethylsulfoxide (DMSO, EMD Millipore Co., anhydrous ≥99.8%). After being fully dissolved, the solution was titrated with methanol (Alfa, 99.9%) until the yellow-orange precipitate did not re-dissolve. Then it was further stirred at 50 °C until a yellow permanent suspension formed. Before the crystal growth, the precursor solution was filtered by a PTFE filter with 0.22 μm pores. Methanol was used as the anti-solvent for slow vapor diffusion and the crystal growth was quenched when the desired morphology was achieved[1]. A similar method was used for the preparation of $MAPbBr_3$ single crystals[2]. The precursor solution (0.45 M) was formed by adding methylammonium bromide (MABr, Dyesol, 98%) and $PbBr_2$ with 1:1 molar ratio into N,N-dimethylformamide (DMF, Aldrich, anhydrous 99.8%). A mixture of dichloromethane ($CH_2Cl_2$, Aldrich, ≥99.5%) and nitromethane (Aldrich, ≥96%) was used as the anti-solvent for $MAPbBr_3$ crystal growth. All solid reactants were dehydrated in a vacuum oven at 150 °C overnight and all solvents were used without further purification.

---
[*] These authors contributed equally.
[†] Current address: LSI, CEA/DRF/lRAMIS, Ecole Polytechnique, CNRS, lnstitut Polytechnique de Paris, F-91128 Palaiseau, France
[‡] To whom correspondence should be addressed. E-mail: xyzhu@columbia.edu

### 1.2. Experimental design

We developed two-dimensional Kerr effect (2D-OKE) spectroscopy based on three major experimental parts: 1.) A Ti:sapphire regenerative amplifier laser system followed by a home-built noncollinear optical parametric amplifier (NOPA); 2.) A pulse shaper based on a transmissive acoustic optic modulator (AOM); and 3.) Balanced detection setup with single-shot data acquisition (DAQ) fully synchronized to the AOM mask sequence. The latter two experimental parts will be explained in separate Method sections.

The ultrashort laser pulses (energy of 0.8 mJ, center wavelength 800 nm, duration of 30 fs, and a repetition rate 10 kHz) were delivered by a Ti:sapphire regenerative amplifier (modified KMLabs Wyvern). These amplified and compressed pulses were split to serve as pump and probe pulses. The major part was used for frequency conversion in a homebuilt NOPA, which is a one stage BBO-based OPA pumped by the second harmonic of the fundamental pulse (center wavelength of 400 nm). The output of the NOPA was optimized for a center wavelength of 550 nm (photon energy of 2.25 eV) and a broad spectrum expanding partly beyond the bandgap of $CsPbBr_3$ (~ 2.35 eV), spanning over 55 nm (230 meV) in full width (see Fig. S3).

The pulse energy incident on the sample was $35 \pm 5$ nJ per pump pulse pair after double pulse generation with the pulse shaper and traversing through a motorized delay stage, focusing optics and polarization optics (broadband waveplate and broadband thin film polarizer). Active dispersion compensation by the pulse shaper was optimized in order to achieve a pulse duration of 30 fs (per single pump pulse), as verified by frequency-resolved optical gating. The probe beam was focused to a diameter of 30 μm under normal incidence and attenuated to a pulse energy of less than 0.2 nJ on the sample. The pump beam was overlapped with the probe focus under a small angle < 5° with respect to the probe beam. The pump beam diameter of 160 μm on the sample was chosen to be significantly larger than the probe beam focus to ensure a homogeneous lateral excitation profile.

For the 1D-OKE excited by a single pump pulse (Figs. 1c,d, S9 and S10), the pulse shaper was bypassed, and the NOPA was optimized for more narrow band spectra with central wavelength of 550 nm (photon energy of 2.25 eV, pulse duration < 30 fs) with spectral widths of < 25 nm.

### 1.3. Pulse and phase shaping

In contrast to previous probe energy resolved OKE,[3] excitation energy resolved 2D-OKE requires a double pulse excitation scheme. To accomplish this, we generated double pulses from the NOPA output pulse by using an AOM based visible pulse shaper (PhaseTech QuickShape Visible)[4]. The phase shaper utilizes the first order diffraction from an acoustic wave, which is used as a phase mask. The phase mask pattern was generated by an arbitrary waveform generator (AWG). The AWG output signal was synchronized to the arrival of the femtosecond laser pulse at the AOM. After each laser shot the next mask of a pre-defined mask sequence was send to the AOM. The phase of each spectral component in the pulse can be modified independently, thereby allowing the creation of double pulses with arbitrary relative phases. Compared to interferometer-based double pulse generation, the use of a pulse shaper exhibits several advantages[4].

First, compared to a mechanical delay stage the pump-pump delay $\tau$ can be controlled more precisely, without mechanical jitter, and much faster on a shot-to-shot basis. This enables a *rapid scan* approach to avoid laser drifts within one full pump-pump delay scan. Even with phase cycling, where every pump-pump delay step consists of 4 different AOM phase masks (see below), in our experiments, a typical pump-pump delay mask sequence of 904 sequences (226 delays) was acquired in 90.4 ms due to single shot detection. Within this time frame (<100 ms) thermal fluctuations of the NOPA and regenerative amplifier are negligible.

Second, the possibility of *phase cycling* is a unique advantage of using a pulse shaper. Due to the collinear propagation and parallel polarization of the two pump pulses $E_1^p$ and $E_2^p$, it is important to separate the nonlinear 2D signals resulting from a coherent interaction of $E_1^p$ and $E_2^p$ from linear and single-pump signals (e.g. transient absorption, scattering or OKE caused be $E_1^p$ and $E_2^p$ separately)[4]. This was achieved by phase cycling, which is implemented by subtracting two signals $S(\tau, \Delta\phi = 0) - S(\tau, \Delta\phi = \pi)$ at the same pump-pump delay $\tau$, where $\Delta\phi$ is the phase difference between the to pump pulses. We used the following 4-sequence phase cycling scheme, which is common in 2D spectroscopy[4]:

$$\bar{S} = S(\phi_1 = 0, \phi_2 = 0) - S(\phi_1 = 0, \phi_2 = \pi) + S(\phi_1 = \pi, \phi_2 = \pi) - S(\phi_1 = \pi, \phi_2 = 0).$$

Additionally, we added a *rotating frame* with frequency $\omega_F$ by an additional phase factor $e^{-i\tau\omega_F}$, which reduced the number of sampling points and increased the frequency resolution in the frequency window of interest[4,5]. Both *rotating frame* and *phase cycling* was automatically implemented by the commercial phase shaper control software (PhaseTech QuickControl).

Importantly, we carefully excluded artifacts related to phase cycling and rotating frame by performing a single pulse excitation (1D) experiment for every 2D-OKE data set. In the single pulse excitation case, we use exactly the same optical path through the phase shaper to account for dispersion and divergence effects. The only difference is the AOM mask sequence, which is simply used as a 5 kHz chopper (masks "1" and "0"). No phase cycling or rotating frame is applied in this chopper mode. The 1D pump-probe delay $t$ is introduced by a conventional mechanical delay stage. As required, for all 2D-OKE data sets, the $E_{\text{ex}}$-integrated transient birefringence signal agrees with the single pulse excitation (1D-OKE) signal (see Figs. 2 and 3).

### 1.4. 2D Data analysis and experimental parameters

To extract the excitation frequency dependence of the observed transient birefringence, we perform a Fourier transform (FT) of the 2D data set along the pump-pump delay $\tau$ as commonly employed in Fourier transform infrared (FTIR) spectroscopy and 2D electronic spectroscopy: Our collinear pump-pump geometry based on a pulse shaper allows for simultaneous phase matching of rephasing and non-rephasing electronic coherences and therefore adds up the corresponding signal contributions[6]. We obtain the oscillatory birefringence as a function of photon energy by taking the real part of the FT with respect to the pump-pump delay $\tau$, as shown in Figs. 2a,c and 3a,c. The perfect agreement of the excitation energy integrated 2D-OKE and the equivalent single pulse excitation experiment (1D-OKE), e.g. in Fig. 2a (top trace), fully validates the consistency and absence of systematic artifacts of the here presented 2D-OKE methodology (projection slice theorem) [6].

We can now treat each horizontal line cut of constant excitation energy $E_{ex}$ as a third order nonlinear polarization response as a function of pump-probe delay $t$. To obtain the OKE frequency spectrum, we thus perform a second FT along the pump-probe axis $t$ and take its absolute value (= amplitude) as we would do for every oscillatory spectrum. Accordingly, this procedure results in the 2D frequency maps showing the correlation between excitation energy and OKE modulation frequency in Figs. 2b,d and 3b,d.

In order to gain optimized energy resolution along the excitation energy axis, experimentally we set the rotating frame (e.g. to 500 THz; 2.07 eV), such that the observation window is centered around the excitation spectrum (e.g. 550 THz; 2.27 eV), see e.g. Fig. S3a,b. The 100 THz width of the accessible frequency window is set by the pump-pump delay step size of 5 fs. The frequency resolution is determined by the maximum pump-pump delay, which was set to 2000 fs.

### 1.5. Balanced detection

In order to detect small changes in the anisotropic transmission, resulting from transient birefringence, we implemented a balanced detection scheme[7]. Our balanced detection enables the observations of very small (~$10^{-4}$) changes in the ellipticity or linear polarization of the ultrashort probe pulse (pulse duration of 30 fs, center wavelength 800 nm) while discriminating against isotropic or unpolarized transmission changes. Therefore, isotropic pump-induced changes such as transient absorption or scattering effects were highly suppressed and did not contribute to the pump probe signal. Further, the phase cycling of the two pump pulses discriminates against incoherent effects[4].

The balanced detection consists of broadband waveplate and Wollaston prism followed by a focusing lens and two identical fast Si-photodiodes. The Wollaston prism spatially separates two orthogonal components of the probe beam; one component parallel and one component perpendicular to the pump pulse polarization. The photo currents corresponding to these two projections are separately single-shot resolved detected on the two identical photodiodes. For small polarization changes the relative difference between signal $\Delta I/I_0 = (I_1 - I_2)/(I_1 + I_2) = \sin(\Gamma)$ scales linear with the phase shift $\Gamma$ between the two electric field polarization components. Depending on the waveplate, this corresponds either to a change in ellipticity or change in linear polarization (polarization rotation angle) for a quarter-wave plate (QWP) or half-wave plate (HWP), respectively. The single shot difference signal $\Delta I(t, \tau, \phi_1, \phi_2)$ was digitized and acquired at a 10kHz rate in conjunction with the corresponding pump-pump-pulse sequence. Here $t$ is the pump-probe delay, $\tau$ is the pump-pump delay, and $\phi_1, \phi_2$ are the phases of the pump pulse sequence.

### 1.6. Kerr propagation model

We simulate the induced 3$^{rd}$ order nonlinear polarization $P_j^{(3)}(t,z)$ according to the general four-wave mixing (FWM) equation (1) in the main paper. The spatial and temporal evolution of all 3 interacting fields are calculated on a time-space grid with a time resolution of $\Delta t' = 10$ fs and one dimensional spatial slices of $\Delta z = 250\ nm$ and 500 nm ($z_i = 0 \ldots d$) for the CsPbBr$_3$ and MAPbBr$_3$ sample, respectively. The coarse spatial resolution (for the sake of computational efficiency) is responsible for the minor phase artifacts in Figs. 4b,c and e,f. The pump-pump and pump-probe delay steps $\Delta \tau$ and $\Delta t$, respectively, are chosen similar to the experimental parameters: $\Delta \tau = 7$ fs ($t = 0 \ldots 2$ ps) and $\Delta t = 15$ fs ($t = -0.2 \ldots 8.5$ ps). We assume that the fields are not depleted by the nonlinear process[8] and that the polarization components $P_j^{(3)}(t,z)$ emit a nonlinear signal field $\delta E^{NL}(t,z)$ at every slice $z$, which then co-propagates with the probe field. The transmitted probe field $E^{pr}$ and integrated nonlinear field $E^{NL}$ is then projected on two polarization detection directions given by a half-wave plate and a Wollaston prism via Jones calculus. The difference of two projected intensities finally delivers the simulated 2D-OKE signal $S(t,\tau)$. A detailed description of this calculation is given in Ref.[9].

To model the 3$^{rd}$ order nonlinear polarization dominated by the instantaneous electronic response, all $\chi^{(3)}$ components were assumed to be independent of frequency $\nu$, which corresponds to a $\delta(t-t')\,\delta(t'-t'')\delta(t''-t''')$ temporal response function. For normal incidence on the (101) crystal surface, the orthorhombic space group *Pnma* of CsPbBr$_3$ permits Kerr signals from $\chi_{xxxx}^{(3)}, \chi_{yyyy}^{(3)}, \chi_{xxyy}^{(3)} = \chi_{xyyx}^{(3)} = \chi_{xyxy}^{(3)}$ and $\chi_{yyxx}^{(3)} = \chi_{yxxy}^{(3)} = \chi_{yxyx}^{(3)}$ (Ref.[1]). The cubic MAPbBr$_3$ at room temperature has space group *Pm3m* and allows for coupling via $\chi_{xxxx}^{(3)} = \chi_{yyyy}^{(3)}$ and $\chi_{xxyy}^{(3)} = \chi_{xyyx}^{(3)} = \chi_{xyxy}^{(3)} = \chi_{yyxx}^{(3)} = \chi_{yxxy}^{(3)} = \chi_{yxyx}^{(3)}$. All allowed tensor elements were assumed to be of same magnitude (in arb. u.).

*Fit to experimental data*

To fit the Kerr propagation simulation to our experimental data and to obtain the quantitative fit parameters shown in Fig. 5, we use a fitting and parametrization procedure detailed in Ref.[9] In short:

First, $t_{1,-}$ is extracted from the phase-discontinuity of experimental energy-time signal (see Fig. S15; results shown as blue diamonds in Figs. 2a,c). $t_{1,-}$ corresponds to the specific pump-probe delay, when the probe pulse cannot catch up with the fast pump components within the sample thickness $d$ (see Fig. 1b). From this we can derive the group index mismatch $\Delta n_g(\nu)$ via Eq. (2) of the main text. The period of the oscillating mixing terms coupling to off-diagonal tensor elements, e.g. $\chi^{(3)}_{xxyy} E_x^p(t',z)\ E_y^{*p}(t',z)$ in Fig. 4d, is governed by the pump-birefringence $\Delta n_0(\nu)$. By fitting our model both to $t_{1,-}$ (blue diamonds in Fig. 2a,c) and to the oscillatory 2D-OKE main branches in Fig. 4f (via FWM phase mismatch $\Delta k^\pm$)[9], we extract consistent values for $\Delta n_0(\nu)$ and $\Delta n_g(\nu)$. Assuming polarization at an angle of $\phi = 0°$ with respect to one of the crystal axes was found to reproduce the experimental data very well (see Fig. 4d-f).

For the 92 K results in Figs. 4 and 5, the refractive and group index at the probe wavelength along the fast axis was set to 1.88 and 2.04, respectively, with a birefringence of 0.033. For the 300 K results, the probe refractive and group index was set to 1.97 and 2.04, respectively, with a birefringence of 0.019. The fast axis pump refractive index $n_x$ at 2.1 eV was assumed to be 1.97 and 2.04 for 92 K and 300 K, respectively, while the dispersion at higher frequencies $n_0(\nu)$ and birefringence $\Delta n_0(\nu)$ is defined by the fitting procedure detailed in Ref.[9].

### 1.7. Time-Frequency analysis

We conduct time-frequency analysis of 1D-OKE using a short-time Fourier transform (STFT), which is available in MATLAB as a standard function. In STFT, a transient is split into windows of constant width, and a fast Fourier transform (FFT) is applied to each window. By specifying window width and overlap, we achieve a time-resolved frequency analysis of the 1D-OKE frequency evolution. In Fig. 1d, the window size was 60 data points and the window overlap was set to 59 data points.

## 2. Extended data analysis

### 2.1. Fluence dependence

As the long data acquisition time of the 2D-OKE experiments does not allow for a reliable fluence dependence measurement series, we investigated both the initial oscillatory high frequency response (region 1) and the consecutively low frequency oscillations (region 2) as a function of pump fluence in a 1D-OKE experiment. The fluence dependence at room temperature, Fig. S10, clearly unveils the linear dependence of both regions on the excitation fluence. This confirms a third order nonlinear polarization as source of the transient birefringence signal and is thus consistent with an OKE response.

In a scenario where the oscillatory OKE response results from coherent lattice oscillations, the only phonon-phonon coupling mechanism suitable to this fluence dependence is based on a linear-square coupling term and a parametric down-conversion type of coupling[10,11]. We can exclude this scenario by a coupled anharmonic oscillator model and by calculating the impact of photodoping or strain in the next sections.

### 2.2. Coupled oscillator model excluding parametric phonon decay

In this section, we present a model of coherently coupled phonons to exclude to phononic origin of the OKE oscillations at 5.3 and 2.65 THz as shown in Fig. 1c,d. We describe the coupling of a strongly-damped high-frequency fully-symmetric $A_\mathrm{g}$ mode to a weakly-damped low-frequency mode of arbitrary symmetry. Our model is fully consistent with a previously described approach for the parametric coupling of the anharmonic $A_\mathrm{g}$ mode in bismuth[10,11]. The harmonic part of the potential energy of the two phonons, $V_\mathrm{h}$, is given by

$$V_\mathrm{h} = \frac{\Omega_1^2}{2} Q_1^2 + \frac{\Omega_2^2}{2} Q_2^2$$

where $Q_1$ and $Q_2$ are the normal mode coordinates (or amplitudes) of the initial 5.3 THz and the target mode given in $\mathrm{\AA}\sqrt{\mathrm{amu}}$, where amu is the atomic mass unit. $\Omega_1$ and $\Omega_2$ are the respective eigenfrequencies given in $2\pi$ THz.

Anharmonic coupling between different phonons is allowed, when their product contains the fully symmetric representation of the system, which, for orthorhombic $CsPbBr_3$, is the $A_g$ representation. For a coherently-excited $A_g$ mode, this leaves nonlinear coupling terms of the type

$$V_{nl} = cQ_1Q_2^2$$

in cubic order of the phonon amplitude, where $Q_1 \equiv Q_{A_g}$, and $c$ is the linear-quadratic coupling coefficient given in $eV/(Å\sqrt{amu})^3$. This is fundamentally different from the nonlinear phonon couplings that have been intensively studied in recent years in the context of ionic Raman scattering, in which a coherently-excited infrared (IR)-active phonon scatters by a Raman (R)-active phonon. This type of interaction is described by a quadratic-linear $Q_{IR}^2 Q_R$s (Refs. [12,13]) or trilinear $Q_{IR,1}Q_{IR,2}Q_R$ (Ref. [14]) coupling and cannot be used to reproduce the temporal signal obtained in this work.

Here, we model the hypothetical excitation of the initial mode via a stimulated Raman scattering process, which can be described by the potential

$$V_R = \varepsilon_0 R Q_1 E^2(t)$$

where $\varepsilon_0$ is the static dielectric constant, $R$ is the frequency-dependent Raman tensor of the $A_g$ mode, and $E(t)$ is the electric field of the incoming laser pulse. The pump pulse $E(t)$ can be modeled as $E(t) = E_0 \exp\{-(t-t_0)^2/[2(\tau/\sqrt{8\ln 2})^2]\} \cos(\omega_0 t + \phi_{CEP})$, where $\tau$ is the full width at half maximum pulse duration, $\omega_0$ is the center frequency of the laser pulse, and the carrier-envelope phase $\phi_{CEP}$ can be chosen to be zero without loss of generality[15]. The resulting system of coupled differential equations is

$$\ddot{Q}_1 + \kappa_1 \dot{Q}_1 + \Omega_1^2 Q_1 = \varepsilon_0 R E^2(t) + cQ_2^2 \quad (S1)$$
$$\ddot{Q}_2 + \kappa_2 \dot{Q}_2 + (\Omega_2^2 + cQ_1)Q_2 = 0 \quad (S2)$$

Here, $\kappa_1$ and $\kappa_2$ are the damping coefficients of the initial and the target mode, respectively. As the initial mode is excited via impulsive excitation, its amplitude depends on the square of the electric field, which is proportional to the fluence $F$: $Q_1 \propto E^2 \propto F$. The $cQ_2^2$ term in Eq. (S1) is the back-action of the coupled target mode, which can be assumed to be negligible due to the strong damping of the initial mode. The $cQ_1$ term in Eq. (S2) dynamically modifies the frequency of the secondary ($Q_2$) mode, and for a resonance condition $\Omega_1 = 2\Omega_2$, the secondary mode gets excited through parametric amplification[10,16]. The amplitude $Q_2$ is linearly dependent on the amplitude of the parametric drive $cQ_1$, and therefore we would expect a linear dependence on the fluence $Q_2 \propto Q_1 \propto F$ consistent with our experimental fluence dependence (Fig. S10 and S11b). We list the parameters that we used for Eqs. (S1) and (S2) in table ST1. Note that, in order to reproduce a similar 1D-OKE signal through nonlinear phonon coupling (see simulation results in Fig. S11), the value of the coupling coefficient that we have to assume here ($\approx$ eV/(Å$\sqrt{\text{amu}}$)$^3$) is two to four orders of magnitude higher than common ground state values found in bulk dielectrics ($\approx$ meV/(Å$\sqrt{\text{amu}}$)$^3$) (Ref. [13]). Such strong anharmonic coupling would be unprecedented and reminiscent of a macroscopic phase transition. We therefore exclude anharmonically coupled phonons as source for the observed oscillatory OKE signal in the orthorhombic phase in both CsPbBr$_3$ and MAPbBr$_3$.

| | |
|---|---|
| $\dfrac{\Omega_1}{2\pi}$ | 5.3 THz |
| $\dfrac{\Omega_2}{2\pi}$ | 2.65 THz |
| $\kappa_1$ | 5.3 THz |
| $\kappa_2$ | 0.75 THz |
| $R$ | 100 Å$^2$/$\sqrt{\text{amu}}$ |
| $c$ | 17 eV/(Å$^2\sqrt{\text{amu}}$)$^3$ |
| $E_0$ | 10 MV/cm |
| $\omega$ | $\hat{=}$ 550 nm |
| $\tau$ | 10 fs |

**Table ST1**. Parameters used in the evaluation of Eqs. (S1) and (S2).

### 2.3. Impact of photodoping and strain on lattice modes

We carried out first principles simulations on both $CsPbBr_3$ and $MAPbBr_3$ to gain insight into the effect of photodoping and structural deformation/stress on the perovskite phonon modes.

We started by considering the perovskites in their orthorhombic phase, considering their neutral and positively charged state, the latter corresponding to the presence of a photogenerated hole. For these systems we calculated the relaxed atomic positions and cell parameters, see results in Fig. S12. As one may notice, for both $CsPbBr_3$ and $MAPbBr_3$ addition of a charge hole to the cell system leads to a contraction of lattice parameters (and associated bond lengths), consistent with previous results[17]. The effect of added electrons should be even smaller, based on previous analyses, so we did not consider it here. The measured lattice contraction refers to that one could expect at very high charge density, as we here consider the extreme case of one positive charge per orthorhombic unit cell (4 formula units).

We then investigate the effect of this lattice contraction on the calculated phonon modes. The simulated IR spectra of both $CsPbBr_3$ and $MAPbBr_3$ are reported for the neutral case in Fig. S13. While the phonon spectrum of $CsPbBr_3$ has a highest accessible frequency of 146 cm$^{-1}$ (4.38 THz), the spectrum of $MAPbBr_3$ extends up to ~3300 cm$^{-1}$ (98.9 THz) due to modes of the organic cations[18]. Here we focus on the low energy region common to both perovskites. The spectra are dominated by Br-Pb-Br bending and Pb-Br stretching modes, which are coupled in $MAPbBr_3$ to librational modes of the methylammonium cation. By inserting a positive charge into the system, the frequencies shift in all cases to higher energies. As an example, the highest $CsPbBr_3$ phonon mode at 146 cm$^{-1}$ shifts to 155 cm$^{-1}$ (4.65 THz). Similarly, for $MAPbBr_3$ the highest mode related to the inorganic framework shifts from 181 to 188 cm$^{-1}$ (5.63 THz). See Table ST2 for more details.

**Table ST2**. Calculated phonon frequencies (cm$^{-1}$) $CsPbBr_3$ and $MAPbBr_3$ in their neutral and positive charge state.

| CsPbBr$_3$ | | MAPbBr$_3$ | |
|---|---|---|---|
| NEU | POS | NEU | POS |
| 11,16 | 20,48 | 20,81 | 18,34 |
| 17,66 | 22,87 | 23,62 | 21,18 |
| 21,66 | 23,04 | 26,4 | 24,51 |
| 23,02 | 24,03 | 28,46 | 25,22 |
| 25,24 | 24,72 | 28,86 | 25,69 |
| 26,31 | 25,02 | 29,04 | 27,46 |
| 27,44 | 26,12 | 31,01 | 30,53 |
| 27,65 | 28,14 | 32,4 | 32,34 |
| 28,17 | 31,55 | 33,54 | 33,9 |
| 29,06 | 32,6 | 36,27 | 36,2 |
| 29,78 | 32,93 | 39,96 | 39,67 |
| 30,13 | 34,02 | 40,15 | 40,26 |
| 32,11 | 34,94 | 41,23 | 43 |
| 32,3 | 37,2 | 44,23 | 44,41 |
| 33,7 | 39,94 | 45,46 | 47,86 |
| 34,73 | 41,77 | 45,68 | 48,9 |
| 35,96 | 45,4 | 47,22 | 49 |
| 36,77 | 46,82 | 47,99 | 52,3 |
| 39,85 | 47,94 | 48,42 | 53,15 |
| 40,41 | 49,26 | 52,2 | 53,95 |
| 40,44 | 49,46 | 53,18 | 56,53 |
| 43,69 | 51,44 | 55,54 | 56,73 |
| 44,21 | 51,86 | 56,72 | 59,03 |
| 44,57 | 54,75 | 57,05 | 59,53 |
| 45,48 | 55,64 | 57,9 | 59,96 |
| 45,57 | 57,33 | 59,83 | 60,38 |
| 51,53 | 59,82 | 64,66 | 60,76 |
| 52,1 | 61,31 | 71,21 | 65,94 |
| 55,81 | 61,41 | 73,91 | 72,05 |
| 56,68 | 65,8 | 75,85 | 75,72 |
| 57,41 | 69,89 | 76,47 | 84,19 |
| 58,51 | 71,48 | 78,38 | 88,74 |
| 60,31 | 71,76 | 79,87 | 89,46 |
| 60,36 | 74,48 | 81,53 | 95,54 |
| 61,8 | 76,66 | 82 | 96,28 |
| 62,34 | 76,85 | 85,02 | 97,14 |
| 62,9 | 81,91 | 87,44 | 98,39 |
| 63,17 | 88,48 | 88,6 | 99,25 |
| 63,57 | 88,87 | 90,13 | 100,35 |
| 63,62 | 90,11 | 91,1 | 101,4 |
| 65,81 | 91,96 | 92,65 | 103,71 |
| 68,57 | 96,22 | 94,06 | 104,74 |
| 70,6 | 98,07 | 97,7 | 105,97 |
| 71,04 | 102,81 | 100,6 | 111,92 |
| 75,96 | 103,33 | 102,51 | 113,47 |
| 79,41 | 103,86 | 109,17 | 119,55 |
| 79,49 | 104,86 | 109,38 | 119,91 |
| 81,6 | 119,5 | 112,68 | 120,96 |
| 83,14 | 119,94 | 115,85 | 121,3 |
| 83,31 | 120,75 | 119,09 | 123,49 |
| 91,2 | 121,15 | 121,27 | 124,96 |
| 112,76 | 135,35 | 125,42 | 136,45 |
| 113,35 | 145,72 | 126,4 | 136,59 |
| 114,32 | 146,05 | 127,18 | 139,15 |
| 137,32 | 150,2 | 129,01 | 140,22 |
| 141,28 | 152,13 | 129,28 | 142,47 |
| 145,85 | 154,88 | 129,7 | 147,19 |
| | | 133,25 | 149,61 |
| | | 141,1 | 149,84 |
| | | 150,21 | 152,66 |
| | | 151,81 | 155,94 |
| | | 152,67 | 156,87 |
| | | 154,15 | 158,08 |
| | | 155,5 | 160,46 |
| | | 167,69 | 172,84 |
| | | 173,07 | 179,72 |
| | | 176,91 | 182,34 |
| | | 179,96 | 186,23 |
| | | 180,96 | 188,38 |
| | | 300,81 | 297,32 |
| | | 303,62 | 297,47 |
| | | 303,64 | 299,91 |
| | | 304,88 | 300,71 |
| | | 903,9 | 901,67 |
| | | 904,16 | 902,5 |
| | | 904,81 | 902,97 |
| | | 905,12 | 906,35 |
| | | 910,48 | 907,69 |
| | | 912,74 | 909,29 |
| | | 913,99 | 910,07 |
| | | 914,52 | 910,25 |
| | | 1004,49 | 1012,94 |
| | | 1005,11 | 1013,39 |
| | | 1005,31 | 1013,58 |
| | | 1005,52 | 1013,78 |
| | | 1237,69 | 1236,72 |
| | | 1238,35 | 1237,61 |
| | | 1238,95 | 1238,59 |
| | | 1239,11 | 1239,05 |
| | | 1246,35 | 1245,04 |
| | | 1246,54 | 1245,55 |
| | | 1246,74 | 1246,33 |
| | | 1247,39 | 1247,43 |
| | | 1390,53 | 1393,48 |
| | | 1391,83 | 1393,75 |
| | | 1392,16 | 1394 |
| | | 1392,66 | 1394,54 |
| | | 1438,42 | 1435,55 |
| | | 1438,86 | 1436,47 |
| | | 1439,54 | 1437,64 |
| | | 1440,68 | 1439,22 |
| | | 1441,67 | 1439,53 |
| | | 1442,59 | 1440,27 |
| | | 1444,21 | 1441,11 |
| | | 1444,64 | 1441,52 |
| | | 1474,27 | 1474,25 |
| | | 1476,55 | 1476,61 |
| | | 1479,47 | 1479,08 |
| | | 1481,76 | 1480,77 |
| | | 1550,96 | 1556 |
| | | 1552,12 | 1557,1 |
| | | 1552,7 | 1557,57 |
| | | 1553,18 | 1558,48 |
| | | 1578,49 | 1579,14 |
| | | 1581,32 | 1582,33 |
| | | 1586,32 | 1587,4 |
| | | 1587,8 | 1588,79 |
| | | 2972,56 | 2980,78 |
| | | 2973,04 | 2981,05 |
| | | 2973,1 | 2981,6 |
| | | 2973,82 | 2981,89 |
| | | 3068,03 | 3077,03 |
| | | 3068,34 | 3077,14 |
| | | 3069,49 | 3078,89 |
| | | 3070,02 | 3079,68 |
| | | 3071,92 | 3079,99 |
| | | 3072,53 | 3080,35 |
| | | 3077,33 | 3087,22 |
| | | 3078,02 | 3088,37 |
| | | 3079,83 | 3090,36 |
| | | 3080,07 | 3090,36 |
| | | 3081,52 | 3090,6 |
| | | 3082,49 | 3091,19 |
| | | 3098,15 | 3099,15 |
| | | 3098,78 | 3100,03 |
| | | 3115,63 | 3117,32 |
| | | 3118,45 | 3119,47 |
| | | 3190,23 | 3247,08 |
| | | 3191,98 | 3250,87 |
| | | 3195,24 | 3252,25 |
| | | 3199,11 | 3256,44 |

These shifts are small considering the experimental observations and are likely not consistent with a photodoping mechanism being responsible for the observed high-frequency oscillatory Kerr response.

We thus investigated whether stress and structural deformation could induce a shift towards higher frequencies. We used in this case a simple cubic model of $CsPbBr_3$ and we compressed one Pb-Br bond recalculating the associated vibration frequencies and energy variation. As one may notice from Fig. S14 to have a 50% highest frequency increase (light dashed lines) one needs to significantly compress the Pb-Br bond with an energy penalty of ~0.2 eV. Clearly, this situation is highly unlikely as this energy is comparable to the calculated formation enthalpy of typical lead-halide perovskites[19]. We can therefore dismiss both dynamic strain or photodoping effects as sources for large phonon frequency shifts.

*Computational Details*

The orthorhombic APbBr$_3$ perovskites were simulated with a 1x1x1 unit cell containing 4 formula units (20 and 48 atoms). All calculations have been carried out using the Quantum Espresso program package[20] along with the PBE functional[21]. Electron−ion interactions were described by scalar relativistic (SR) ultrasoft pseudopotentials with electrons from N, and C 2s2p, H 1s, Pb 6s6p5d, Cs 4s4p4d and Br 4s4p electrons explicitly included in the calculations. A 2x2x2 k-point mesh for sampling the Brillouin zone was used[22]. Calculations were performed using plane wave cutoffs of 25 and 200 Ry for expansion of the wave function and density, respectively, while for variable cell calculations and for calculations in the presence of the field 50 and 400 Ry cutoffs are used. For the simulation of the structural compression we adopt a cubic CsPbBr3 unit cell, we carried out a variable cell geometry optimization using a cutoffs of 50/400 Ry along with a 4x4x4 k-points grid. DFT-D2 dispersion interactions have been included in the calculation[23]. From the relaxed cell we constrained one Pb-Br at different bond lengths and we relax all the other atoms keeping the relaxed cell parameter fixed. On these optimized geometries a phonon calculation has been performed at the same level of theory.

x[24–26]

# 3. Extended data figures

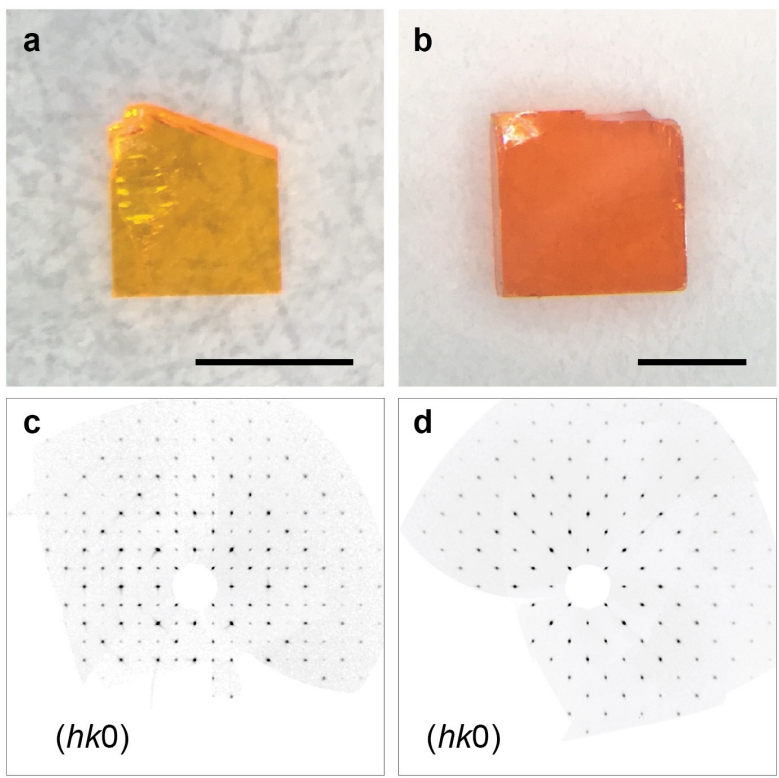

**Fig. S1| Sample characterization.** Optical images (**a**,**b**) and diffraction patterns (**c**,**d**) of $CsPbBr_3$ and $MAPbBr_3$ single crystals. Scale bars are 1 mm. The single-crystal X-ray diffraction (SCXRD) was collected at room-temperature and 180 K for $CsPbBr_3$ and $MAPbBr_3$, respectively. The clear optical morphology and single crystalline nature indicate the high quality of the crystals.

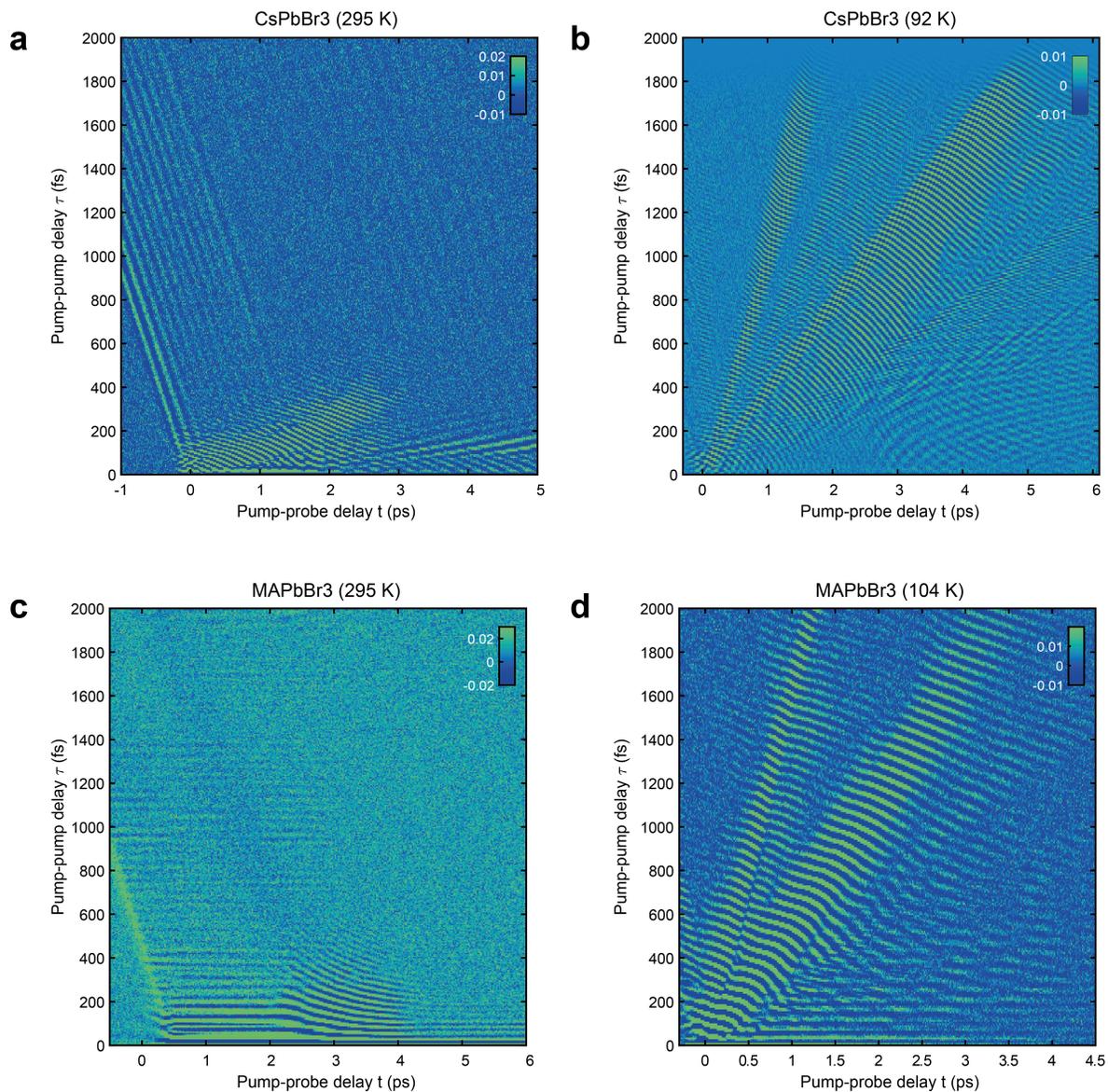

**Fig. S2| Time-time domain data.** Time-time domain raw data from 2D-OKE. Displayed signals $S(t,\tau)$ are retrieved after 4-sequence phase cycling scheme (see Methods). Therefore, each data pixel consists of 4 pump-pump-probe experiments with distinct pump-pump phase. Before time-zero (negative slope) straight lines are single-pump OKE contributions from phase-cycling imperfections due to small laser drifts during hours of data acquisition. All data correspond to the 2D-OKE signals shown in the main paper: **a,** CsPBr$_3$ at 295 K. **b,** CsPBr$_3$ at 92 K (Fourier window function superimposed for $\tau > 1800$ fs). **c,** MAPBr$_3$ at 295 K. **d,** MAPBr$_3$ at 104 K.
S17

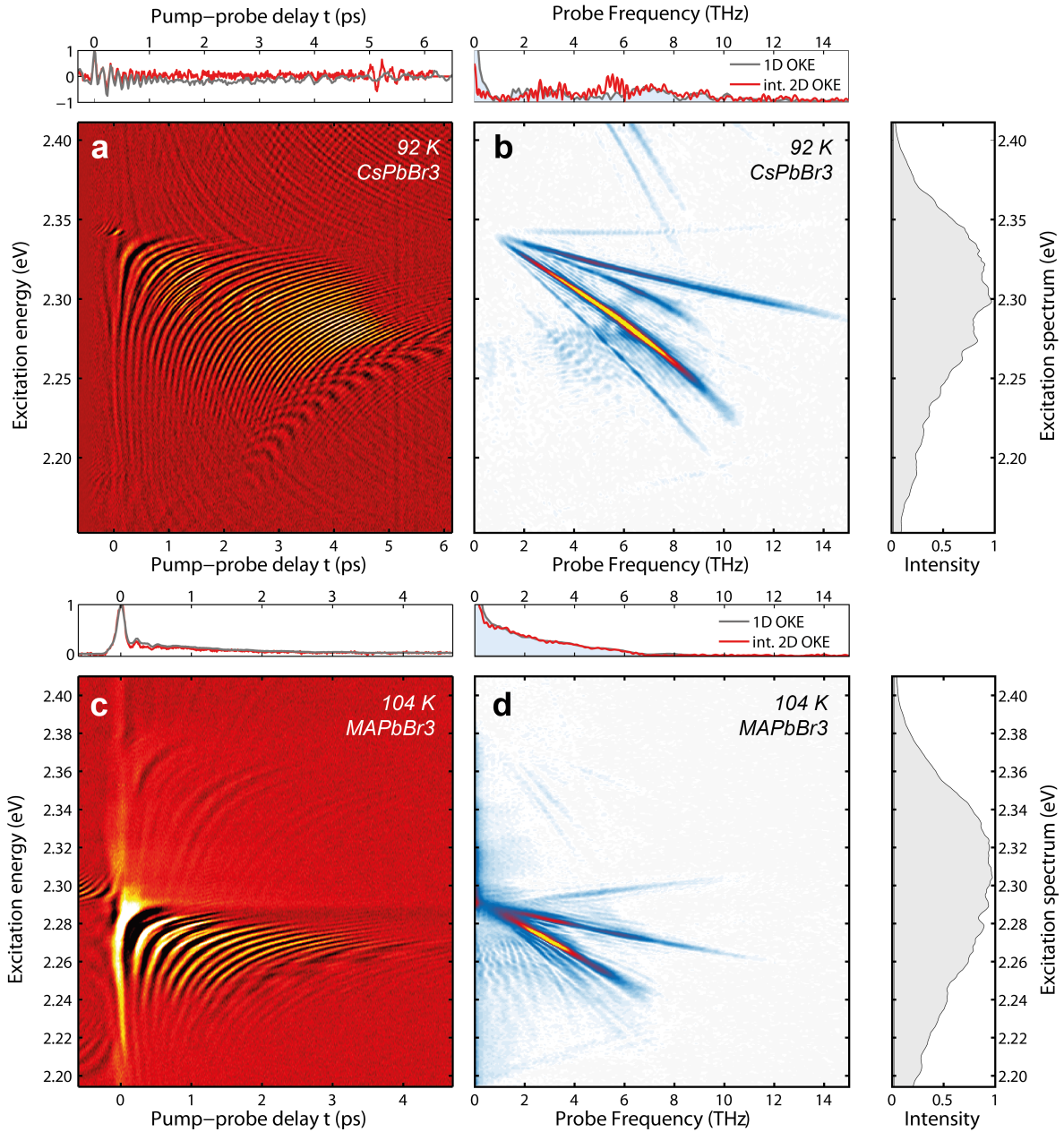

**Fig. S3| Low temperature 2D-OKE incl. excitation energy spectrum. a**, **b,** Low temperature 2D-OKE signal of CsPbBr$_3$ at 92 K in energy-time and energy-frequency domain, respectively. The upper bound of the oscillatory response at ~2.35 eV indicates the bandgap edge and its steep absorption feature; consistent with temperature dependent photoluminescence and absorption studies (Ref. 24). The pump-pulse spectrum (right-hand side) clearly shows that this upper bound of the 2D-OKE is not related to a lack of pump photons with energies above the bandpag. **c,d,** Same as a,b, but for MAPbBr$_3$ at 104 K. Clearly, the 2D-OKE feature precisely pinpoints the optical band gap at 2.29 eV (Ref. 24) despite the much broader excitation energy spectrum (right-hand side).

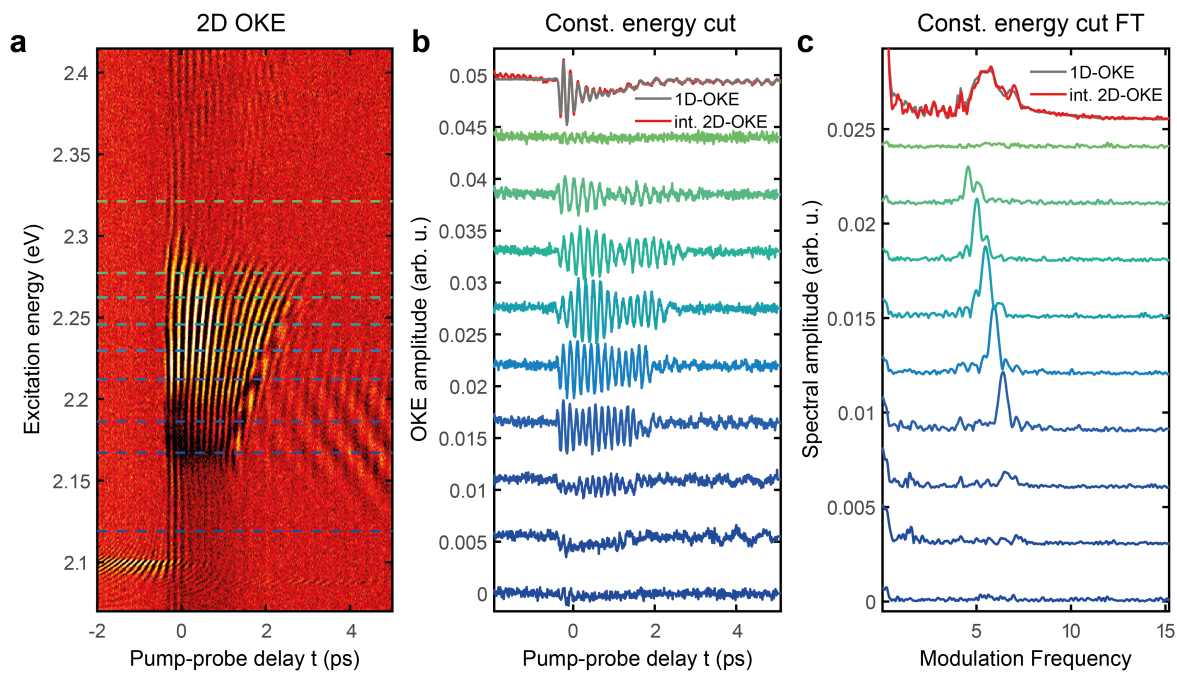

**Fig. S4| Line cuts for specific excitation energies (CsPbBr$_3$). a, b,** Time-frequency domain 2D-OKE at 295 K (as in Fig. 2a) and corresponding line cuts for fixed excitation energies, respectively. The time domain traces in **b** are vertically offset for clarity. The excitation energy-integrated 2D-OKE (red line) agrees well with the single pulse excitation (grey line). Small modulation for $t < 0$ results from the replica at 2.1 eV, which is caused by imperfect phase cycling in conjunction with the rotating frame. **c**, Corresponding Fourier transforms of the line cuts in **b**. Already in these few line cuts a significant frequency shift is prominent. The dominant high-frequency feature sits on top of a smaller low-frequency modulation.

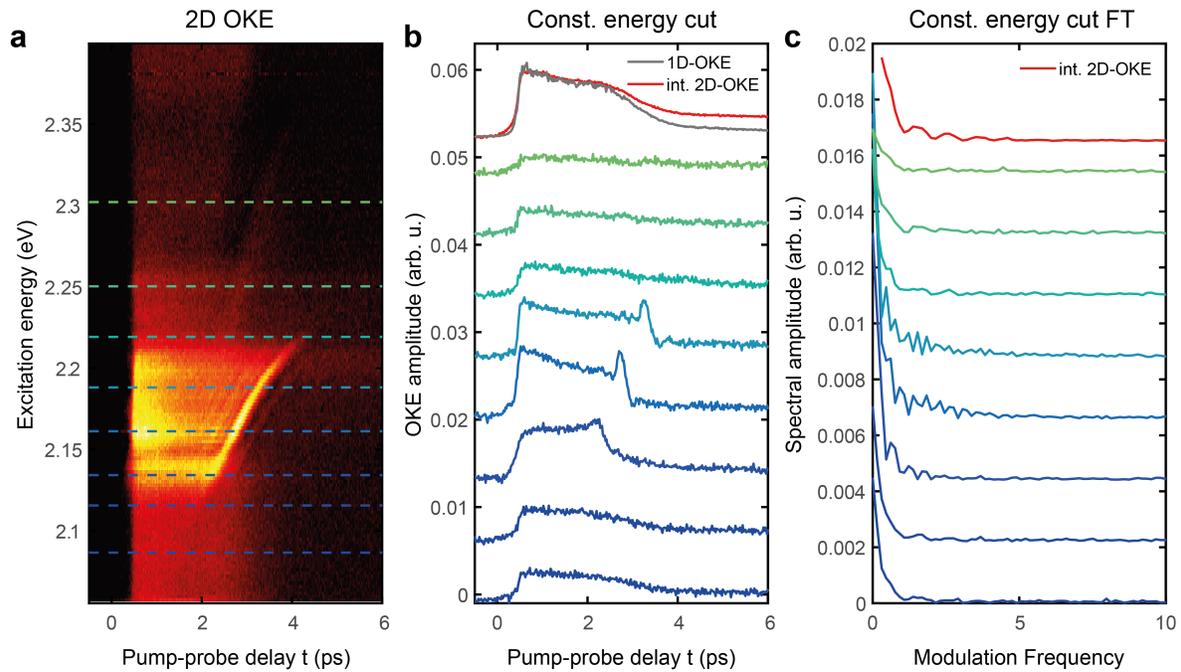

**Fig. S5| Line cuts for specific excitation energies (MAPbBr$_3$). a, b,** Time-frequency domain 2D-OKE at 295 K (as in Fig. 3a) and corresponding line cuts for fixed excitation energies, respectively. The time domain traces in **b** are vertically offset for clarity. The window-like Kerr feature is fully consistent with previous below-gap 1D-OKE studies. The spike at the end of the window-like feature results from the dispersed pump-pulse mixing with its reflection at the backside of the sample. **c,** Corresponding Fourier transforms of the line cuts in **b**. The spike leads to small ripples observed for frequencies up to 5 THz; also visible in Fig. 3b.

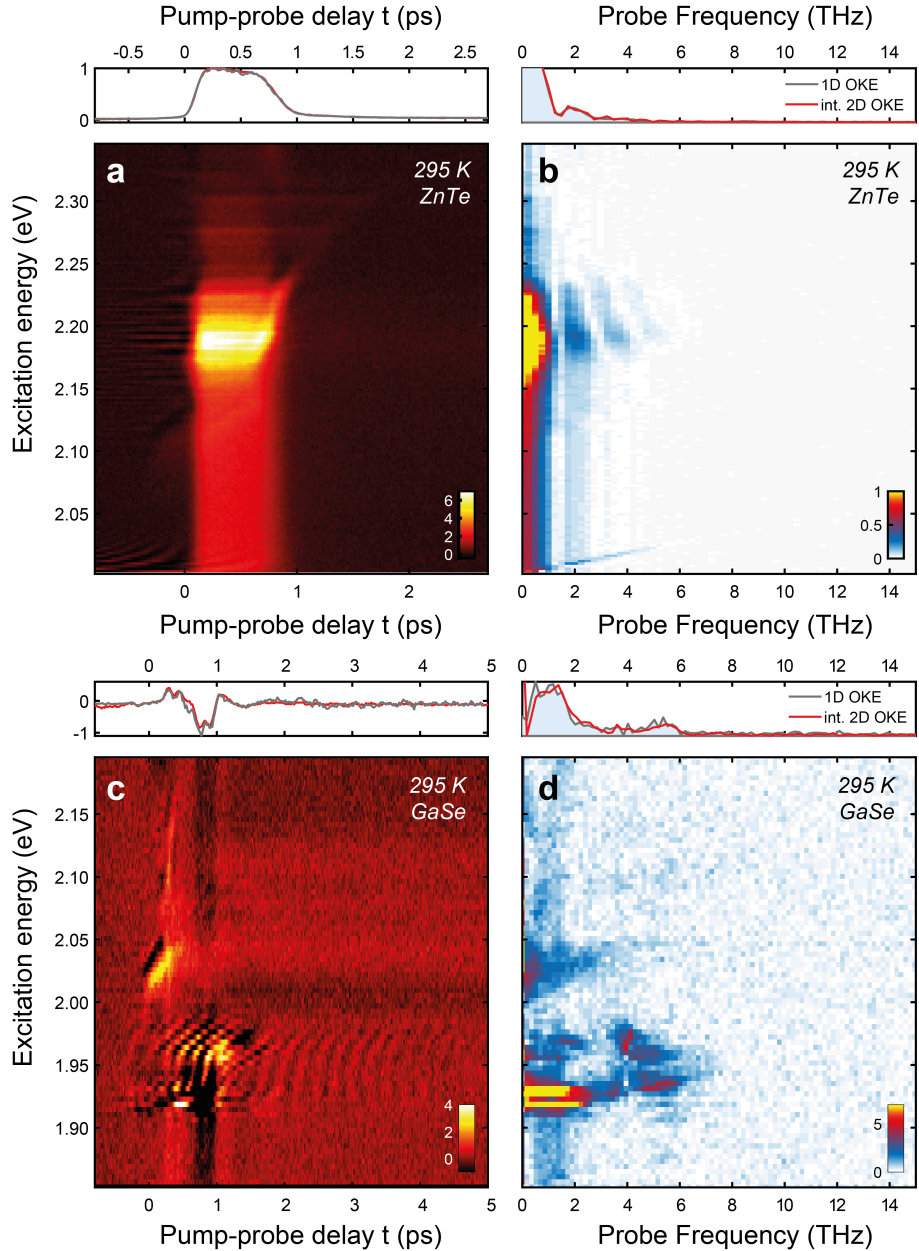

**Fig. S6| Comparative 2D-OKE in ZnTe and GaSe at 295 K. a**, 2D-OKE of the isotropic (cubic) semiconductor ZnTe with a bandgap of 2.25 eV. We clearly find a similar non-oscillatory Kerr-response as in the cubic phase MAPbBr$_3$. The curvature of the ZnTe feature is less pronounced, consistent with the lower refractive index dispersions in ZnTe (Ref. 23). **b**, 2D-OKE frequency spectrum of **a**. **c**, 2D-OKE of the birefringence (hexagonal) semiconductor GaSe with bandgap at 2.03 eV (Ref. 23). The GaSe birefringence of $\Delta n_0 \approx 0.2$ at 1.95 eV (Ref. 25) is one order of magnitude higher than $\Delta n_0$ of LHPs close to the band gap as derived in this work. Nevertheless, only under specific phase matching angles, an oscillatory signal similar to orthorhombic phase perovskites was observed. This comparison highlights the robustness and the strength of the oscillatory signal from polarization dressed light propagation in LHPs. **d,** 2D-OKE frequency spectrum of **c**. The features between 3-6 THz show mild similarities to the correlation branches in the 2D frequency maps of orthorhombic LHPs.

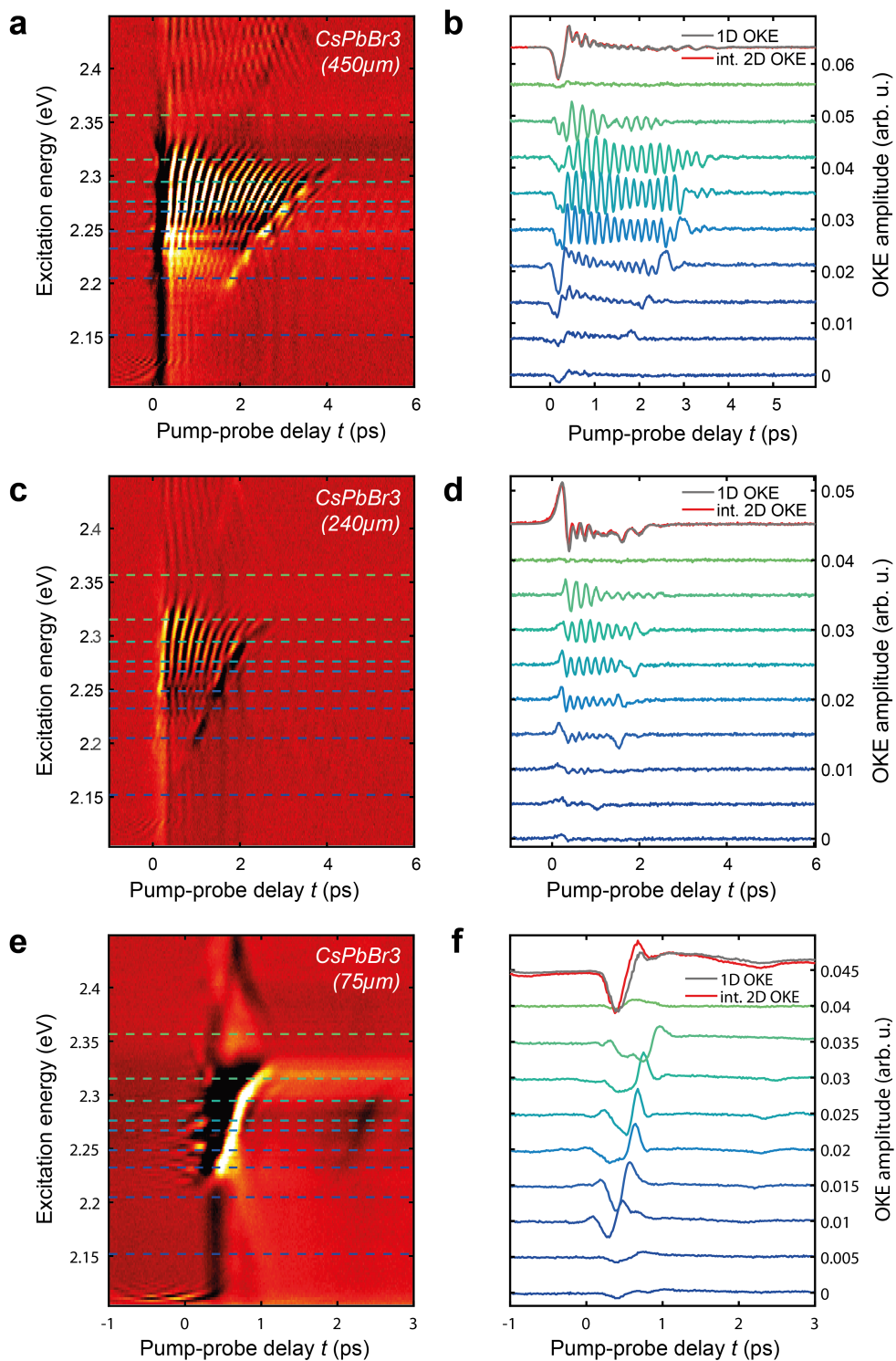

**Fig. S7| Thickness dependence. a, c, e,** 2D-OKE signal of CsPbBr$_3$ at 295K for different thicknesses of 450μm, 240μm, 75μm, respectively. **b, d, f,** corresponding constant excitation energy cuts to visualize the temporal width of the main oscillatory signal. Signal width as a function of thickness for $E_{ex} = 2.29$ eV is given in Fig. S8.

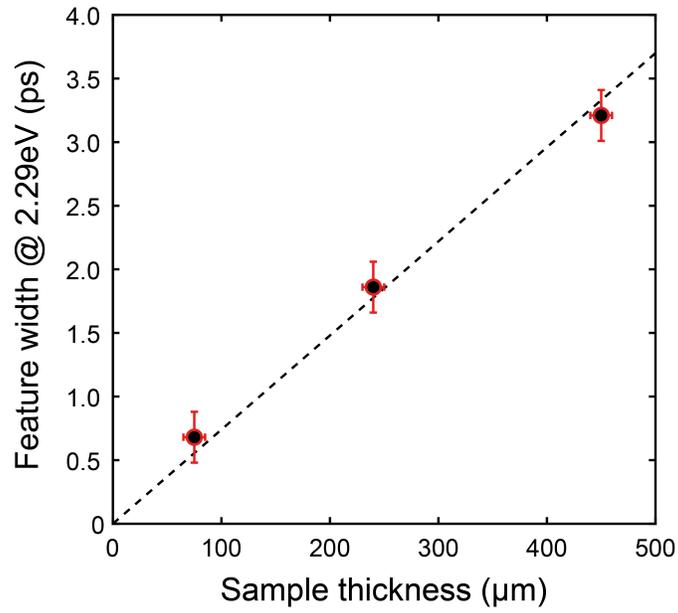

**Fig. S8| Thickness dependence for 2.29 eV excitation energy. a**, 2D-OKE width as a function of single crystal CsPbBr$_3$ thickness at 295 K. For an arbitrary cut a constant excitation energy of 2.29 eV, the duration of the main 2D-OKE feature scales linearly with the thickness. This demonstrates that 2D-OKE traces the nonlinear polarization by light propagation along the full propagation path through the sample.

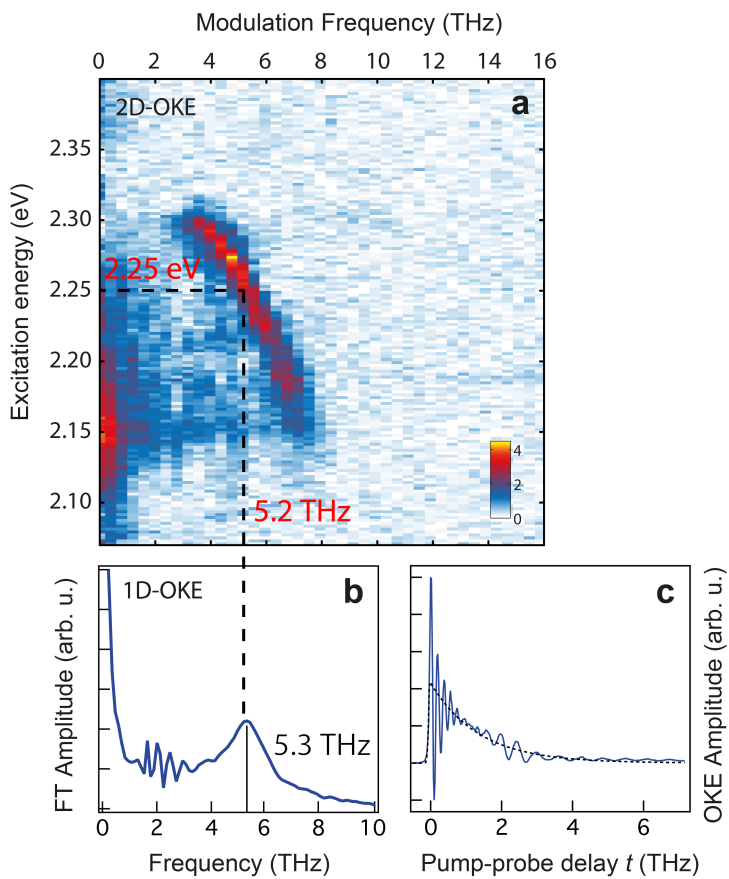

**Fig. S9 | Consistency of 1D-OKE and 2D-OKE. a,** Pump-energy resolved Fourier transform of 2D-OKE on $CsPbBr_3$ at 295K. As the pump excitation energy decreases from the bandgap, the OKE modulation frequency increases with excitation energy from 3.5 THz to 7.0 THz. The dotted line indicates that at an excitation energy of 2.25 eV the corresponding dominating OKE frequency is 5.2 THz. **b, c** The Fourier transform and 1D-OKE trace of $CsPbBr_3$ with a narrowband excitation centered at 2.25 eV, respectively. The initial high-frequency response in c corresponds to the broad peak centered at 5.3 THz in **b**. Note that **a,** and **b,** share the same frequency axis. The close match between the dotted line (5.2 THz) and the solid line (5.3 THz) emphasizes the consistency of 1D- and 2D-OKE.

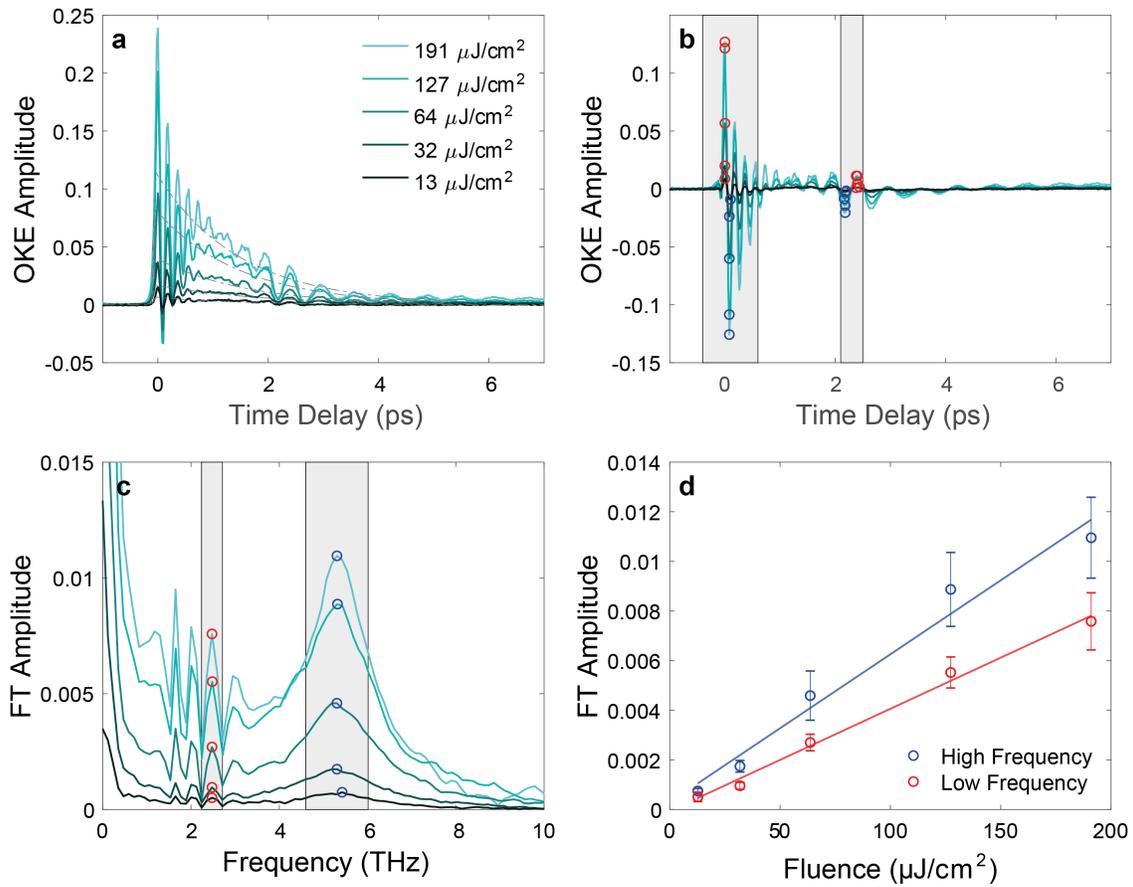

**Fig. S10 | Fluence-dependent 1D-OKE in CsPbBr$_3$ at 295K. a,** The raw data taken with pump fluences of 191, 127, 64, 32, and 13 µJ/cm$^2$. The incoherent contribution is shown with gray-broken traces (single-exponential decay). **b,** The coherent contribution to the OKE trace, taken by subtracting the incoherent part in **a** from the respective raw trace. The gray shaded regions indicate areas in which the high-frequency and low-frequency phonon contributions are separated. The red markers indicate the maximum in the gray shaded regions, the blue markers indicate the minimum in the same regions. **c,** The Fourier transform (FT) of the raw data shown in **a**. The blue and red markers indicate the maximum FT amplitude of the respective gray shaded regions. **d,** The maxima of the peaks in **c** plotted against pump fluence. As shown, both the higher frequency and lower frequency features scale linearly with fluence. This confirms a third order nonlinear polarization as source of the transient birefringence signal.

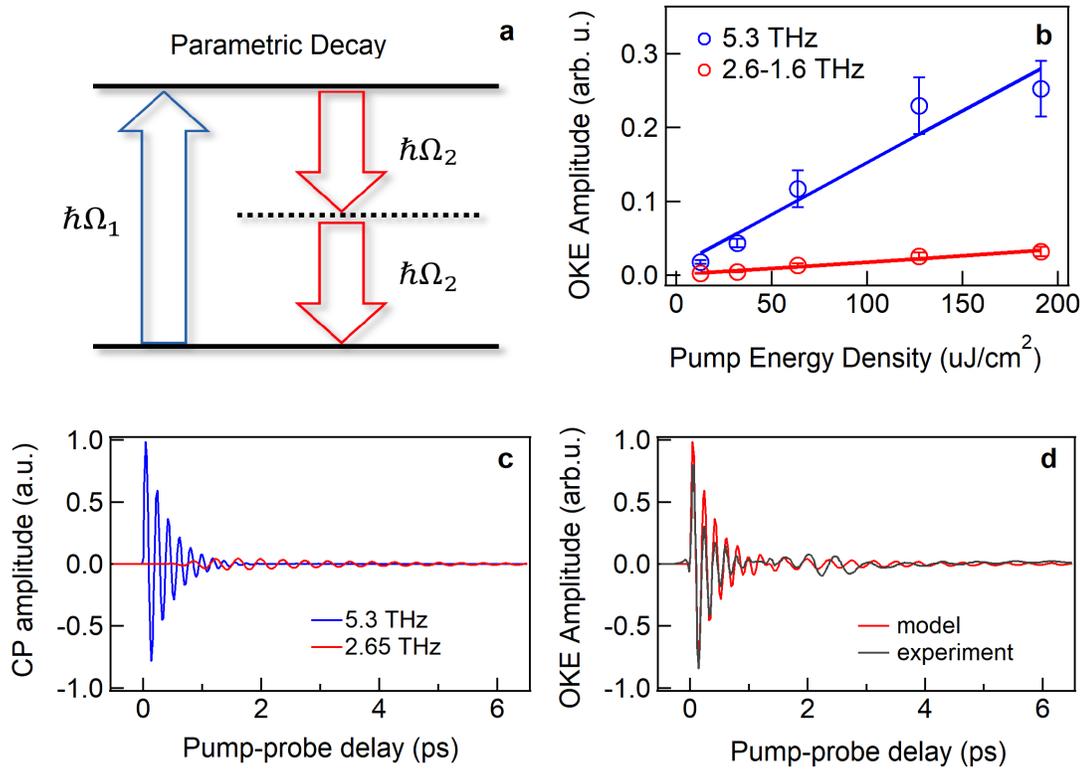

**Fig. S11 | Model simulation of coherent phonon-phonon coupling and comparison to 1D-OKE signal in CsPbBr$_3$ at 295 K. a,** Schematic illustration of parametric decay; **b,** Excitation fluence dependence of the oscillatory OKE signal extracted from time domain data in Fig. S10b. Blue-circles: the 5.3 THz feature (region 1); red circles: the 2.6-1.6 THz feature (region 2). The lines are linear fits. **c,** Simulation results from Eqs. (S1) and (S2) of the time-evolution of the coherent coupling of a 5.3 THz (blue) to a 2.65 THz (red) mode. **d,** The simulated coherent phonon trace (red) can be fitted to rough agreement with experimental result from Fig. 1c (black), but this would require an unprecedented high anharmonic coupling constant $c$ in Eqs. (S1) and (S2).

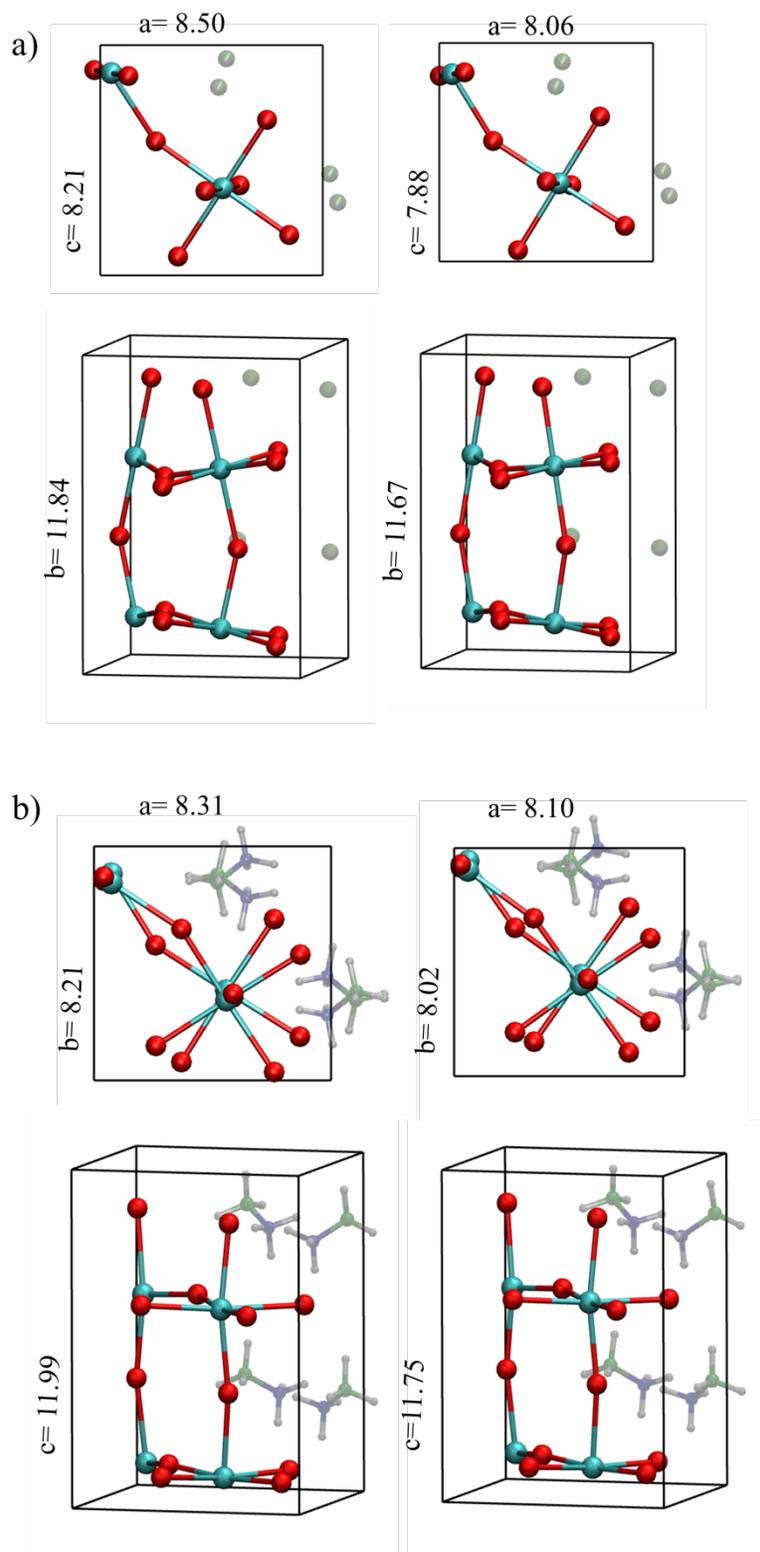

**Fig. S12 | Optimized structures and lattice parameters.** a) CsPbBr$_3$ and b) MAPbBr$_3$ in their neutral (left) and positive (right) charge states along with calculated cell parameters (Å). Different views are reported in upper (along *c* axis) and lower (along the *a* axis) panels.

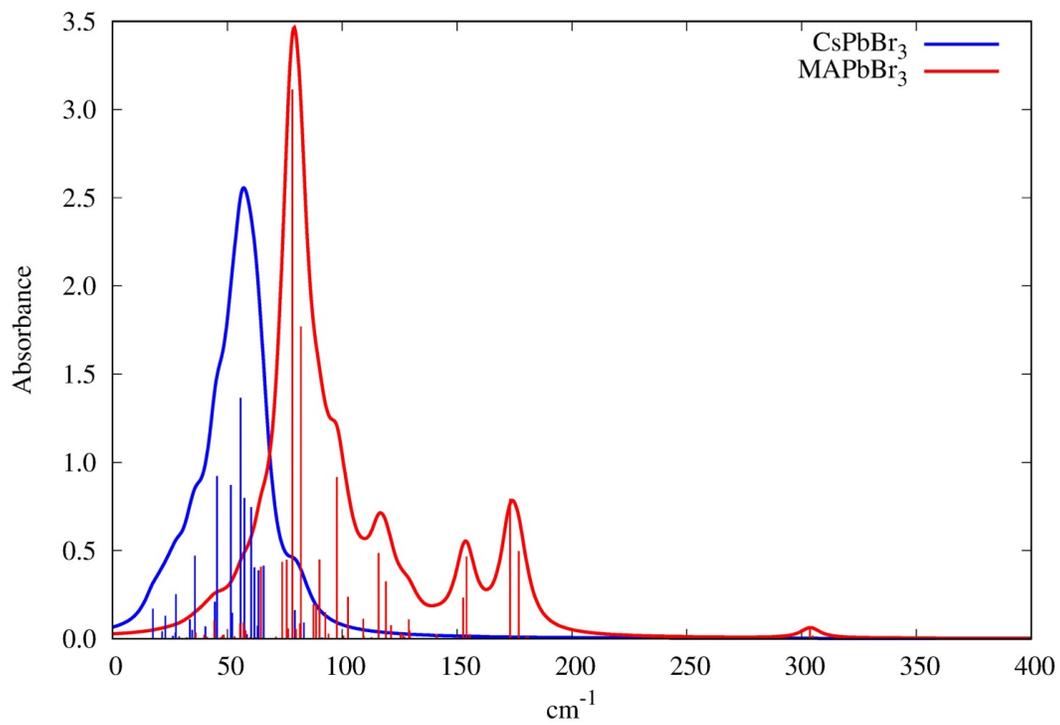

**Fig. S13 | Simulated IR spectra.** For CsPbBr$_3$ and MAPbBr$_3$ perovskites in the low frequency range.

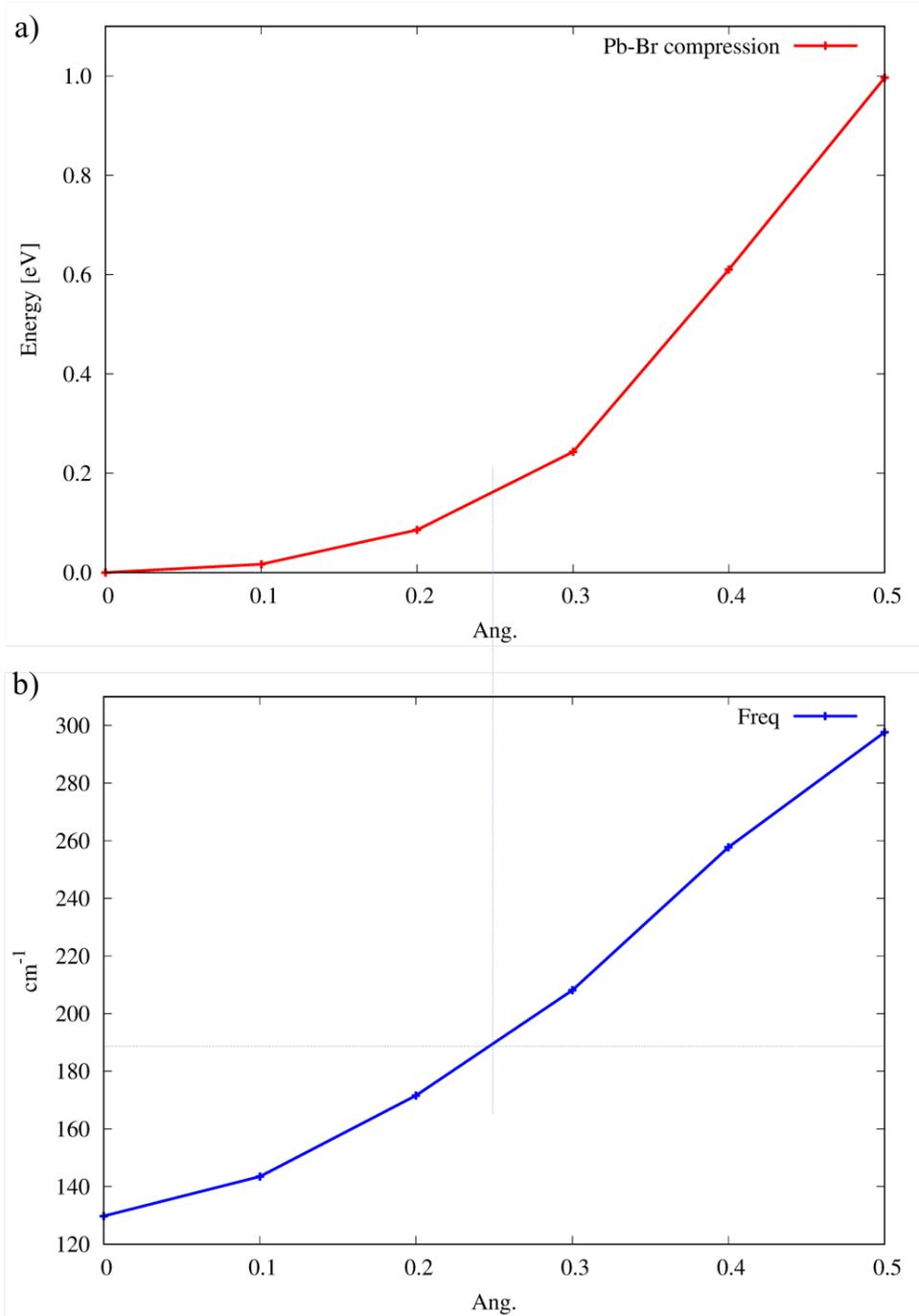

**Fig. S14** | Energy variation (a, eV) and calculated highest phonon frequency (b, cm$^{-1}$) for cubic CsPbBr$_3$ subjected to compression of a single Pb-Br bond (Å).

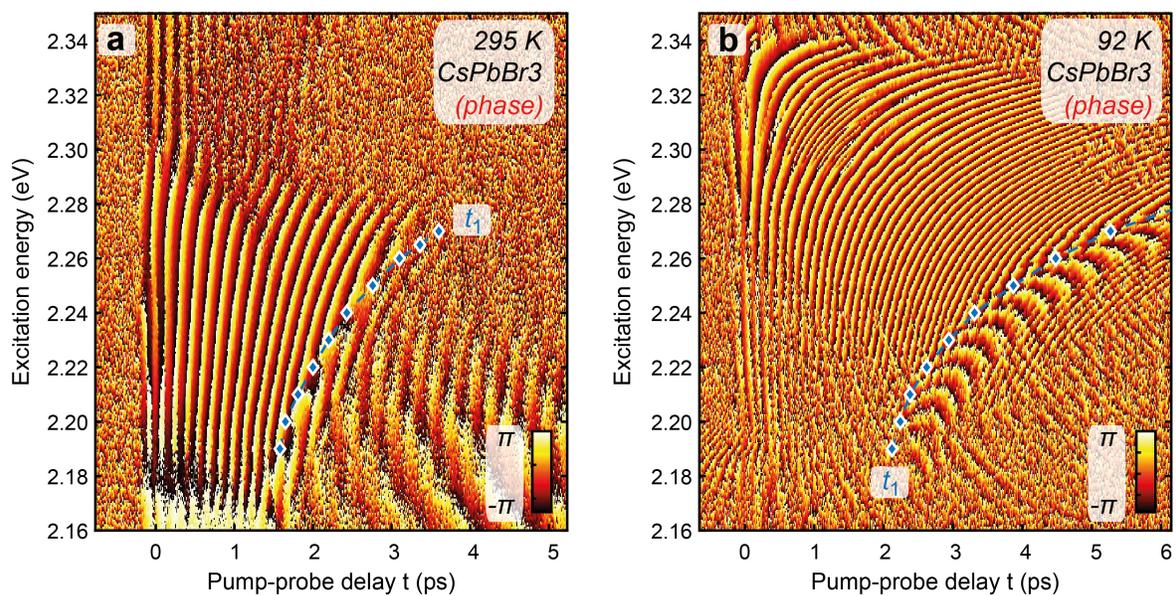

**Fig. S15 | Phase discontinuity.** The phase discontinuity of the 2D-OKE signals of Fig. 2) is used to determine the excitation frequency dependent stop $t_{1,-}(\nu)$ of the high-frequency feature (region 1) for CsPbBr3 **a,** at 295 K and **b,** at 92 K.